\documentclass[12pt,english]{article}
\pdfoutput=1
\usepackage[T1]{fontenc}
\usepackage[utf8]{inputenc}
\usepackage{subfigure,lmodern, amsmath,amssymb, graphicx, pifont, adjustbox, bm, xcolor}
\usepackage{amsfonts}
\usepackage{geometry}
\usepackage{comment}
\usepackage{mathtools}
\usepackage{float}
\usepackage{slashed}
\usepackage{cite}
\usepackage{ragged2e}
\usepackage{etoolbox}
\usepackage{array}

\apptocmd{\thebibliography}{\justifying\setlength{\leftskip}{7.4mm}}{}{}
\makeatletter\g@addto@macro\bfseries{\boldmath}\makeatother

\usepackage{stackengine}
\usepackage{esint}
\usepackage[unicode=true,pdfusetitle,
 bookmarks=true,bookmarksnumbered=false,bookmarksopen=false,
 breaklinks=false,pdfborder={0 0 1},backref=false,colorlinks=true]
 {hyperref}
\hypersetup{pdfauthor={Clifford Cheung and Grant N. Remmen},
 citecolor=black,linkcolor=black,urlcolor=black}

\newcommand{\appendixref}[1]{\hyperref[#1]{appendix~\ref{#1}}}
\def\equationautorefname~#1\null{eq.\,(#1)\null}
\usepackage{breakurl}
\usepackage[hang,flushmargin]{footmisc} 
\allowdisplaybreaks

\usepackage{changepage}

\newcommand{\be}{\begin{equation}}
\newcommand{\ee}{\end{equation}}
\newcommand{\bea}{\begin{eqnarray}}
\newcommand{\eea}{\end{eqnarray}}

\newcommand{\eq}[2]{\be\begin{aligned}#1 \label{#2}\end{aligned}\ee}

\newcommand{\Fig}[1]{Fig.~\ref{#1}}

\newcommand{\Eq}[1]{Eq.~(\ref{#1})}

\newcommand{\Sec}[1]{Sec.~\ref{#1}}

\newcolumntype{P}[1]{>{\centering\arraybackslash}p{#1}}

\begin{document}

\interfootnotelinepenalty=10000
\baselineskip=18pt
\hfill

\thispagestyle{empty}
\begin{center}
{\LARGE \bf
Veneziano Variations:\\ \bigskip
\Large How Unique are String Amplitudes?
}\\
\bigskip\vspace{1cm}
\begin{center}{\large Clifford Cheung${}^{a}$ and Grant N. Remmen${}^{b,c}$}\end{center}
\vspace{7mm}
{
\it ${}^a$Walter Burke Institute for Theoretical Physics,\\[-1mm]
California Institute of Technology, Pasadena, California 91125\\[1.5mm]
${}^b$Kavli Institute for Theoretical Physics,\\[-1mm] University of California, Santa Barbara, CA 93106\\[1.5mm]
${}^c$Department of Physics,\\[-1mm] University of California, Santa Barbara, CA 93106}
\let\thefootnote\relax\footnote{e-mail: 
\url{clifford.cheung@caltech.edu}, 
\url{remmen@kitp.ucsb.edu}}
 \end{center}

\bigskip
\centerline{\large\bf Abstract}
\begin{quote} \small

String theory offers an elegant and concrete realization of how to consistently couple states of arbitrarily high spin.  But how unique is this construction?
In this paper we derive a novel, multi-parameter family of four-point scattering amplitudes exhibiting i)~polynomially bounded high-energy behavior and ii)~exchange of an infinite tower of high-spin modes, albeit with a finite number of states at each resonance.  These amplitudes take an infinite-product form and, depending on parameters, exhibit mass spectra that are either unbounded or bounded, thus corresponding to generalizations of the Veneziano and Coon amplitudes, respectively.  For the bounded case, masses converge to an accumulation point, a peculiar feature seen in the Coon amplitude but more recently understood to arise naturally in string theory~\cite{Maldacena:2022ckr}.  Importantly, our amplitudes contain free parameters allowing for the customization of the slope and offset of the spin-dependence in the Regge trajectory.  
We compute all partial waves for this multi-parameter class of amplitudes and identify unitary regions of parameter space.  For the unbounded case, we apply similar methods to derive new deformations of the Veneziano and Virasoro-Shapiro amplitudes.
\end{quote}
	
\setcounter{footnote}{0}

\setcounter{tocdepth}{2}
\newpage
\tableofcontents

\newpage

\section{Introduction}  \label{sec:intro}

The amplitudes bootstrap exploits the serendipitous fact that many of the theories that describe nature are actually fixed by simple conditions imposed on the kinematic functions that encode scattering.  For example, while gauge theory and gravity can famously be derived from high-minded geometric principles or gedankenexperiments, this is not our only recourse for understanding their origins.  Rather, an alternative path to discovering gauge theory and gravity is to answer a concrete {\it math question} about the S-matrix:  What is the space of on-shell scattering amplitudes for massless spin-one and spin-two particles that are local, Lorentz invariant, and mediate a long-range force?

It is then natural to ask, what is the analogous question in string theory?  In broad terms, string theory furnishes a working example of how gravity and quantum mechanics can be rendered consistent.   At a more technical level, it addresses the problem that high-energy, fixed-angle graviton scattering is ill behaved.  Unitarizing this behavior requires the addition of higher-spin degrees of freedom that must be included with care lest they exacerbate these ultraviolet pathologies.  Thus, string theory offers an explicit answer to the concrete question of how to consistently build an amplitude that exhibits the exchange of higher-spin modes and is sensible at high energies.  In his seminal work \cite{Veneziano:1968yb}, Veneziano constructed precisely such an amplitude,
\be
\begin{aligned}
A_{\rm string}(s,t) & = \frac{\Gamma(-\alpha_0-\alpha' s)\Gamma(-\alpha_0-\alpha' t)}{\Gamma(-2\alpha_0-\alpha' (s+t))}.
\end{aligned}\label{eq:Ven}
\ee  
In this paper, we revisit this line of inquiry, in particular asking to what extent string amplitudes are the unique solutions to this particular math problem.

Concretely, we construct a multi-parameter space of Lorentz invariant, four-particle, perturbative scattering amplitudes $A(s,t)$ that describe an infinite exchange of higher spins while conforming to the following conditions:
\begin{itemize}

\item[i)] {\it Polynomial Boundedness}. The high-energy behavior of the amplitude is polynomially bounded.  If this condition fails, causality and unitarity are generically violated \cite{Froissart:1961ux,Chaichian:1987zt,Camanho:2014apa,
Haring:2022cyf
}.\footnote{Rigorous bounds exist for gapped theories, e.g., the Regge ($A < s^2$) and Froissart ($A<s\log^{D-2} s$) bounds, which apply to fixed momentum transfer and scattering angle, respectively.}

\item[ii)] {\it Finite-Spin Exchange}. Each pole in the amplitude describes the exchange of a finite tower of spins.  When this condition fails, the exchanged state is an infinitely extended object, thus undermining the locality of the theory.

\end{itemize}

 \noindent Famously, string theory provides precisely a function $A(s,t)$ that conforms to these criteria in the form of Eq.~\eqref{eq:Ven}.  In this case, the amplitude vividly encodes the Regge spectrum of an infinite tower of stringy excitations, evenly spaced in mass-squared and with an ever-growing collection of spins. 
 
 Perhaps less appreciated is the existence of {\it yet another} solution to the above constraints, discovered---and then sadly forgotten by virtually all---in the remarkable work of Coon~\cite{Coon}.  The Coon amplitude is a deformation of the Veneziano amplitude labeled by a single parameter~$q$.   The most notable attribute of the Coon amplitude is its hydrogen-like spectrum: a discrete array of mass levels that converge to infinite density at an accumulation point, followed by a branch cut.\footnote{A reasonable worry is that an infinite density of states is inconsistent with statistical mechanics because a thermal ensemble will sample the infinite reservoir of modes near the accumulation point.  Similar logic would imply tension with holographic bounds on degrees of freedom.  However, recall that any such pathology requires that the degenerate states be indistinguishable.  If they are not---for example for the case of the hydrogen atom and some recent string theoretic constructions \cite{Maldacena:2022ckr}---then nothing is awry.} Recently, a number of works \cite{Tourkine, Geiser:2022icl,Chakravarty:2022vrp} have analyzed the Coon amplitude from the perspective of positivity bounds derived from analytic dispersion relations.    By demanding that the spectral density be nonnegative---i.e., no ghosts---those authors mapped out a putative consistent region for $q$.
 
The outline of the present work is as follows.   In \Sec{sec:bootstrap}, we generalize the analysis of Coon to derive a new multi-parameter family of amplitudes that are cyclic invariant on the external legs and conform to the criteria described above.  
These amplitudes exhibit the same mass spectrum of accumulation points as the Coon amplitude, but differ markedly in their distribution of spins at each level.  Interestingly, they also exhibit {\it customizable} Regge trajectories.  We also describe the low- and high-energy limits of these amplitudes.
We emphasize that this multi-parameter class is not an exhaustive solution space but merely offers an existence proof of such deformations.\footnote{In this work we restrict to the case of scalar external states.  As is well known, however, the inclusion of spin is achieved simply by multiplying scalar amplitudes by polarization-dependent prefactors.  Unfortunately, this operation generically worsens the high-energy and Regge behavior of the amplitudes.}

Afterward, in \Sec{sec:positivity} we derive constraints on this family of amplitudes coming from partial wave unitarity.  This enforces the nonnegativity of every coefficient in the partial wave decomposition of the amplitude on each residue.  We derive both analytic and numerical positivity bounds and observe that there is a multi-parameter subspace of putatively consistent amplitudes.

In \Sec{sec:variations} we then derive a new family of amplitudes obeying the above criteria but exhibiting mass spectra that are free from accumulation points and thus unbounded.  We generalize the analysis of \Sec{sec:bootstrap} to this case and present a multi-parameter space of amplitudes exhibiting either cyclic or full permutation invariance on the external legs.  
Unfortunately, this alternative class of theories appears to be disfavored by unitarity constraints.  

Finally, we summarize our results and discuss promising future directions in \Sec{sec:discussion}.

\section{Amplitude Construction}  \label{sec:bootstrap}

\subsection{Product Ansatz}

In this section we construct a family of amplitudes $A(s,t)$ describing the scattering of external scalars subject to the constraints of polynomial boundedness and finite-spin exchange described in \Sec{sec:intro}.   Here we assume cyclic invariance on the external legs, so $A(s,t) = A(t,s)$ is $st$-symmetric, like color-ordered amplitudes in gauge theory and open string theory.

If we view $A(s,t)$ as a tree-level amplitude, then it exhibits simple poles on exchanged resonances and can be written as a rational function of the form
\eq{
A(s,t) = \frac{N(s,t)}{\prod\limits_{n=0}^\infty(s-m^2_n)(t-m^2_n)}.
}{}
Here $m^2_n$ describes an arbitrary spectrum indexed by nonnegative integers $n$, properly ordered such that $m_n^2 < m_{n+1}^2$.  Note that any state degeneracy at a given mass level is accounted for by multiplicity factors in the numerator.

At this juncture, let us call attention to an important subtlety: infinite products such as $\prod\limits_{n=0}^\infty(s-m^2_n)$ are actually {\it ill defined}.  In particular, convergence requires each factor in the product to approach unity at large $n$, which is impossible for all $s$.  To express the amplitude in a convergent form, one needs to make a binary choice of whether the spectrum is unbounded or bounded.  For the latter case, $m^2_{\infty}$ is finite and there is an accumulation point in the spectrum.  In the subsequent analysis we will assume that this is the case.   Note that precisely for this reason, none of our results will contradict those of Ref.~\cite{Caron-Huot:2016icg}, whose analysis argued for the uniqueness of the Veneziano amplitude, but given the explicit assumption that the spectrum is free from accumulation points.

With a spectrum bounded from above and below, it is convenient to define the auxiliary kinematic variables $\sigma,\tau$,
\eq{
s &= (m_0^2 - m_{\infty}^2) \sigma + m_{\infty}^2 \\
t &= (m_0^2 - m_{\infty}^2) \tau + m_{\infty}^2,}{eq:Mandelstam}
which are related to the physical Mandelstam variables $s,t$ by an affine transformation.   Here the range $0\leq \sigma\leq 1$ maps to $m_{\infty}^2 \geq s\geq m_0^2$, which scans through the spectrum of states.  In terms of the auxiliary kinematic variables, the amplitude takes the form
\eq{
A(\sigma,\tau) = \frac{N(\sigma,\tau)}{\prod\limits_{n=0}^\infty\left(1-\frac{f(n)}{\sigma}\right)\left(1-\frac{f(n)}{\tau}\right)},
}{}
which exhibits an infinite tower of simple poles located at
\eq{f(n) = \frac{m^2_n - m^2_\infty}{m^2_0 - m^2_\infty},}{eq:fdef}
so each factor in the product converges to unity for infinite $n$.

It appears exceedingly difficult---if not impossible---to derive the general space of functions $A(\sigma,\tau)$ conforming to the conditions of polynomial boundedness and finite-spin exchange.  For this reason, we now make some highly nontrivial simplifying assumptions about the form of the numerator.  Inspired by the infinite-product form of the Veneziano and Coon amplitudes, we assume a product ansatz for the numerator, so
\eq{
A(\sigma,\tau) = W(\sigma,\tau) \prod_{n=0}^\infty \frac{1 - g(n)\left(\frac{1}{\sigma} + \frac{1}{\tau}\right) - \frac{h(n)}{\sigma\tau}}{\left(1-\frac{f(n)}{\sigma}\right)\left(1-\frac{f(n)}{\tau}\right)}.
}{eq:prod}
Here $f(n)$, $g(n)$, and $h(n)$ are as yet undetermined functions, while $W(\sigma,\tau)$ is a general prefactor function that is analytic in the region $\sigma,\tau\in(0,1)$. 
As with $f(n)$, we assume that $g(n)$ and $h(n)$ vanish in the $n\rightarrow \infty$ limit so that the numerator product converges on its own. 
Note that \Eq{eq:prod} was also the starting point of the original analysis of Coon~\cite{Coon}, who ultimately input additional assumptions that we will relax in the subsequent analysis. 

\subsection{Polynomial Residue Constraint}\label{sec:STC}

Amplitudes that are expressed as {\it products} rather than {\it sums} of propagators generically exhibit the peculiarity of an infinite tower of higher-spin states at a {\it single} resonance.  For example, an amplitude of the form
\eq{
\frac{1}{(s-m^2)(t-m^2)}
}{eq:nonlocal}
has a residue at $s=m^2$ that is an infinite series in $t$, thus implying an infinite-spin tower  at a single mass level.  
Curiously, amplitudes such as this play an important role in extremizing the positivity bounds derived from the EFT-hedron \cite{Caron-Huot:2020cmc,Arkani-Hamed:2020blm}.  More recently, they have been shown to provide alternative unitarizations of four-point graviton scattering~\cite{Huang:2022mdb}.  

Clearly, \Eq{eq:nonlocal} strongly violates the condition of finite-spin exchange described earlier.
Physically, the exchange of an infinite-spin resonance suggests a breakdown of locality, since the object exchanged is unboundedly macroscopic.
This scenario contrasts with the Regge spectrum of string theory, where each mass level supports a finite tower of spinning states.
In the present work, we restrict to a more conservative scenario in which infinite-spin resonances are forbidden, so locality is guaranteed.  Mathematically, the constraint of finite-spin exchange simply implies that the residue on any pole in the $s$ channel is a finite degree polynomial in $t$.  We refer to this as the constraint of {\it polynomial residues}, and we now impose this condition on our ansatz.

Returning to the amplitude ansatz in \Eq{eq:prod}, consider the $s$-channel pole at mass level $k$, corresponding to the limit $\sigma\rightarrow f(k)$.  
For clarity, let us define numerator and denominator factors as $N_n(\sigma,\tau) = 1-g(n)\left(\frac{1}{\sigma}+\frac{1}{\tau}\right) - \frac{h(n)}{\sigma\tau}$ and $D_n(\sigma) = 1-\frac{f(n)}{\sigma}$.
The residue $R_k(\tau)$
at this $k$th pole is then
\be
\begin{aligned}
R_k(\tau)
 &= \lim_{\sigma\rightarrow f(k)}\left(1-\tfrac{f(k)}{\sigma}\right)A(\sigma,\tau) \\&= W(f(k),\tau)\lim_{\sigma\rightarrow f(k)}D_k(\sigma)\prod_{j=0}^\infty [D_j(\sigma)]^{-1} \prod_{n=0}^\infty \frac{N_n(f(k),\tau)}{D_n(\tau)} \\&
= W(f(k),\tau)\prod_{j,\,j\neq k}[D_j(f(k))]^{-1} \frac{\prod_{n=0}^\infty N_n(f(k),\tau)}{\prod_{m=0}^\infty D_m(\tau)} 
\\&= W(f(k),\tau)\prod_{j,\,j\neq k}[D_j(f(k))]^{-1} \frac{\prod_{l=0}^{p(k)-1}N_l(f(k),\tau)\prod_{n=p(k)}^{\infty} N_n(f(k),\tau)}{\prod_{m=0}^\infty D_m(\tau)} \\
& =W(f(k),\tau)\prod_{j,\,j\neq k}[D_j(f(k))]^{-1}\prod_{l=0}^{p(k)-1}N_l(f(k),\tau)\prod_{n=0}^{\infty}\frac{N_{n+p(k)}(f(k),\tau)}{D_n(\tau)},
\end{aligned} \label{eq:shift}
\ee
where in the second-to-last line we have rearranged the product by peeling off the first $p(k)$ terms of the numerator.  
In the final line we have shifted the index of the remaining infinite tower of numerator factors by $p(k)$. 
We emphasize that this rearrangement was possible because of our requirement that the numerator and denominator products in the ansatz in \Eq{eq:prod} separately converge, allowing us to write the products over the numerator and denominator separately in the third line above. 
In the analysis of Coon, it was simply assumed without fanfare that $p(k)=k$, in which case the residue on the $k$th pole exhibits the cancellation of an infinite set of poles except for $k$ leftover factors from the numerator product.  Here we instead generalize $p(k)$ to an arbitrary monotonic function.

The constraint of polynomial residues is the requirement that the $\tau$-dependence cancels in the final infinite product in the last line of \Eq{eq:shift}, which implies that\footnote{Strictly speaking, this constraint does not guarantee that the final answer is a polynomial in $\tau$, but merely a ratio of polynomials. This subtlety will come into  play later.}
\be 
\phi(n,k) =1-g(n+p(k))\left[\frac{1}{f(n)}+\frac{1}{f(k)}\right]-\frac{h(n+p(k))}{f(n)f(k)}=0,\label{eq:STC}
\ee
where for later convenience we define a function $\phi$ whose vanishing enforces the constraint of polynomial residues.
In what follows, let us consider how this requirement constrains the space of functions $p$, $f$, $g$, and $h$.

\subsubsection{Fixing the $p$ Function}

Without loss of generality, let us replace $n$ with an equivalent variable $r=n+p(k)$, in which case the constraint of polynomial residues becomes
\eq{
\phi(r-p(k),k)= 1-g(r)\left[\frac{1}{f(r-p(k))}+\frac{1}{f(k)}\right]-\frac{h(r)}{f(r-p(k))f(k)}=0.
}{}
Solving algebraically for $f(k)$, we obtain
\eq{
f(k) = \frac{f(r-p(k))g(r)	+h(r)}{f(r-p(k))-g(r)}.
}{}
Plugging the left-hand side directly back into the right-hand side, we find
\eq{
f(k) = f(r-p(r-p(k))) .
}{eq:fconstraint}
For integer $k$, well orderedness of the $m_n^2$ and the definition in Eq.~\eqref{eq:fdef} imply that $f(k)$ is invertible, so Eq.~\eqref{eq:fconstraint} implies that $r-k = p(r-p(k))$, which in terms of $n$ yields
\eq{
p(k)-k = p(n)-n.
}{}
For this relation to hold for any $n$ and $k$ implies that $p$ is linear, with unit slope and arbitrary offset,
\eq{
p(k) = k+ \alpha.
}{}
Thus, we have arrived at a new parameter $\alpha$, which was implicitly assumed to be zero in the analysis of Coon.  Since $p(k)$ and $k$ are nonnegative integers in Eq.~\eqref{eq:shift}, we know that $\alpha$ is also a nonnegative integer.

\subsubsection{Fixing the $f, g, h$ Functions}\label{sec:fgh}

The condition of polynomial residues in \Eq{eq:STC} is an immensely stringent constraint on the functions $f$, $g$, and $h$ when they are evaluated at {\it nonnegative integer values} of $n$ and $k$.  However, following similar logic as Ref.~\cite{Coon}, we can identify the space of {\it smooth} functions that satisfy \Eq{eq:STC} for {\it arbitrary real values} of $n$ and $k$.  This isolates a subclass of functions of the real numbers for which \Eq{eq:STC} holds.

Starting from the polynomial residue constraint in \Eq{eq:STC}, we define new functions $\phi^{(i,j)}(n,k)$, obtained by taking from $i$ derivatives with respect to $n$ and $j$ derivatives with respect to $k$.   This quantity must vanish for any value of $n$ and $k$, so in particular we consider
\eq{
\phi^{(0,0)}(n,n) = \phi^{(1,0)}(n,n) =\phi^{(2,0)}(n,n) =\phi^{(3,0)}(n,n)  =\phi^{(4,0)}(n,n) =0.
}{eq:phi_derivs}
The vanishing of the first expression implies that $h(2n+\alpha) = f(n)\left[f(n)-2g(2n+\alpha)\right]$, which automatically sets the second expression to zero.  Plugging $h$ back into the third expression, we find that $g(2n+\alpha) = \left[ f(n) f''(n) - f'(n)^2 \right]/f''(n)$, which also sets the fourth expression to zero. 
By finiteness of $g$, we can further conclude that $f''(n)\neq 0$ unless $f'(n)=0$, which is forbidden by our requirement that $f$ be strictly monotonic to avoid repeated poles. Finally, plugging both $h$ and $g$ back into the fifth expression, we obtain the differential equation
\eq{
f'(n)^2 f''''(n) -4 f'(n)f''(n)f'''(n)+3 f''(n)^3 =0,
}{eq:fdiffeq}
which can be compactly written as
\be
\{ f,n\}= {\rm constant}, \label{eq:Schwarzian}
\ee
where $\{f,n\} = \tfrac{f'''(n)}{f'(n)} - \tfrac{3f''(n)^2}{2f'(n)^2}$ is the Schwarzian derivative of $f$ with respect to $n$.
Without loss of generality, we set the right-hand side of \Eq{eq:Schwarzian} to $-(\log q)^2/2$ for some arbitrary constant $q$.  The solution to this equation is
\eq{
f(n) &= \frac{a+b q^n}{c+d q^n}\\
g(n) &= \frac{bd q^n- ac q^\alpha}{d^2 q^n-c^2 q^\alpha}\\
h(n) &= -\frac{b^2 q^n - a^2 q^\alpha }{d^2 q^n - c^2 q^\alpha},
}{eq:fghsolution_temp}
where $f$ is a M\"obius transformation of $q^n$ parameterized by the constants $a,b,c,d$, which are defined modulo rescaling such that $ad-bc \neq 0$, so the possible spectra span a representation of ${\rm SL}(2,\mathbb{R})$. As expected from the fact that \Eq{eq:fdiffeq} is a fourth-order differential equation, \Eq{eq:fghsolution_temp} describes a four-parameter space of solutions.

As described in \Eq{eq:fdef}, $f(n)$ interpolates between the boundary points $f(0)=1$ and $f(\infty)=0$.
Monotonicity of $f(n)$ implies that $q>0$ and $q \neq 1$.  Meanwhile, since sending $q\rightarrow 1/q$ is equivalent to swapping $a\leftrightarrow b$ and $c\leftrightarrow d$, we take that $q<1$ without loss of generality.  The boundary conditions then fix $a=0$ and $d=b-c$, reducing the four-parameter space of solutions down to a two-parameter expression.  Defining $c$ as $(1+\xi) b$ for later convenience, we obtain 
\be
\begin{aligned}
f(n) &= \frac{q^n}{1+\xi(1-q^n)} \\
g(n) &= \xi h(n) \\
h(n) &= \frac{1}{q^{\alpha-n}(1+\xi)^2 - \xi^2},
\end{aligned} \label{eq:fghsolution}
\ee
which is our general solution to the polynomial residue constraint.

\subsubsection{Fixing the $W$ Function}

By construction, on the residue of the amplitude at $\sigma\rightarrow f(k)$, an infinite tower of $\tau$-dependent factors exactly cancels between the numerator and denominator.  However, as is clear from \Eq{eq:shift}, the residual $\tau$-dependence in the residue yields a polynomial of degree $k+\alpha$ in the variable $1/\tau$.  Thus the highest inverse power of  $\tau$ in the residue at $\sigma\rightarrow f(k)$ is
\eq{
R_k(\tau)
 \propto \tau^{-(k+\alpha)} W(f(k),\tau) +\cdots .
}{}
To maintain locality of the amplitude, any inverse powers of $\tau$ must be cancelled, so the prefactor $W(f(k),\tau)$ has to contain a compensating factor of $\tau^{k+\alpha}$ or even higher powers of $\tau$.  While this function can in principle be a polynomial in $\tau$, we assume for simplicity that it is monomial, so 
\eq{
W(f(k),\tau) = e^{{\cal M}(f(k))} \tau^{{\cal N}(f(k))} ,
}{}
where ${\cal N}(f(k)) \geq k+\alpha$ is an integer controlling the power of $\tau$ and ${\cal M}(f(k))$ is a real number controlling the size of the numerical coefficient.  By definition, $W(\sigma,\tau)=W(\tau,\sigma)$ since this prefactor multiplies an $st$-symmetric function in \Eq{eq:prod}.  This implies that $W(\sigma,\tau) = e^{{\cal M}(\sigma)}\tau^{{\cal N}(\sigma)}=e^{{\cal M}(\tau)}\sigma^{{\cal N}(\tau)}$, or equivalently that
\eq{
{\cal M}(\sigma) + {\cal N}(\sigma)\log\tau =  {\cal M}(\tau) + {\cal N}(\tau)\log\sigma.
}{}
The unique solution to this constraint is ${\cal N}(\sigma) = {\cal Y} +  {\cal X} \log \sigma$ and ${\cal M}(\sigma) = {\cal Z} +  {\cal Y}\log \sigma$, so 
\eq{
W(\sigma,\tau)  = e^{{\cal Z}+{\cal Y}(\log \sigma+\log \tau)+{\cal X} \log\sigma \log \tau}.
}{}
Now, in order to make sure that the function
\eq{
{\cal N}(f(k)) &=
{\cal Y}+ {\cal X}\log\left[\frac{q^k}{1+\xi(1-q^k)} \right]\\
&\overset{\xi=0}{=} {\cal Y}+{\cal X} k \log q = \alpha+\beta +(1+\gamma)k \geq k+\alpha,
}{}
is integer-valued, we henceforth require that $\xi=0$, as inserted in the second line.  
Demanding that the resulting monomial in $\tau$ is of degree at least $k+\alpha$, we then require that ${\cal X}= (1+\gamma)/\log q$ and ${\cal Y}=\alpha+\beta$, where $\beta$ and $\gamma$ are nonnegative integers.  
In conclusion, the final expression for the prefactor is
\eq{
W(\sigma,\tau) = c_{\alpha\beta\gamma}(\sigma\tau)^{\alpha+\beta} q^{(1+\gamma)\frac{\log\sigma}{\log q}\frac{\log\tau}{\log q}},
}{eq:Wpre}
for nonnegative integers $\alpha,\beta,\gamma$.  Here we have recast the overall constant normalization as $e^{\cal Z}=c_{\alpha\beta\gamma}$, which can in principle depend on the other parameters.

Note that the parameter $\beta$ is actually somewhat trivial: unlike $\alpha$ and $\gamma$, it simply multiplies the amplitude by a power of $\sigma\tau$.  Nevertheless, we will see later that this relatively innocuous addition will modify the consistency properties of the amplitudes, in particular regarding partial wave unitarity. 

\subsection{Final Amplitude}

To summarize, after fixing the $f,g,h$, and $W$ functions in our ansatz in \Eq{eq:prod}, we obtain our final expression for the amplitude,
\eq{
A(\sigma,\tau) &=  c_{\alpha\beta\gamma} (\sigma\tau)^{\alpha+\beta} q^{(1+\gamma)\frac{\log \sigma}{\log q}\frac{\log \tau}{\log q}} \prod_{n=0}^\infty \frac{(1-\frac{q^{n-\alpha}}{\sigma\tau})}{\left(1-\frac{q^n}{\sigma}\right)\left(1-\frac{q^n}{\tau}\right)}.
}{eq:Afull} 
Here $\alpha,\beta,\gamma \in \mathbb{N}_0 \in \{0,1,2,\cdots \}$ are nonnegative natural numbers and $0<q<1$, yielding a four-parameter space of amplitudes modulo overall normalization.

\subsubsection{Physical Spectrum}

As described in \Eq{eq:Mandelstam}, the auxiliary kinematic variables $\sigma$ and $\tau$ are related by an affine transformation to the physical Mandelstam variables $s$ and $t$.  Consistent with  \Eq{eq:Mandelstam}, we now define a convenient parameterization that has been employed in previous analyses \cite{Tourkine,Maldacena:2022ckr,Geiser:2022icl,Chakravarty:2022vrp},
\be 
\begin{aligned}
\sigma & =1+(q-1)\left(\frac{s}{\mu^{2}}-\delta\right)\\
\tau & =1+(q-1)\left(\frac{t}{\mu^{2}}-\delta\right).
\end{aligned}\label{eq:sigmatau}
\ee
The variables $\delta$ and $\mu^2$ technically constitute {\it fifth and sixth} physical parameters for our amplitudes.  These quantities did not make an appearance in our earlier analysis because they have no effect on the constraint of polynomial residues and only appear when mapping $\sigma$ and $\tau$ to $s$ and $t$.  Since $\mu^2$ simply controls the overall scale of masses, it is largely irrelevant.  In contrast, $\delta$ sets the offset of the lightest mass from zero and hence will be crucial in our analysis of partial wave unitarity. 

The mapping in \Eq{eq:sigmatau} implies that our amplitude in \Eq{eq:Afull} has the {\it exact same mass spectrum} as the Coon amplitude, 
\be
m_n^2 = \mu^2 \left(\delta + [n]_q\right),\label{eq:spoles}
\ee
where $[n]_q$ is the $q$-deformed integer,
\be
[n]_q = \frac{1-q^n}{1-q} = 1+q+q^2 + \cdots q^{n-1}.\label{eq:nq}
\ee
As in the case of the Coon amplitude, we will assume that the mass-squared of the external states is equal to that of the lowest state in the spectrum, $m_0^2 = \mu^2 \delta$.
Also, note that the $q\rightarrow 1$ limit generates the evenly spaced and unbounded Regge spectrum in masses, $m_n^2\sim n$, of the Veneziano amplitude of string theory.

\begin{figure}[tp]
\begin{center}
\includegraphics[width=11cm]{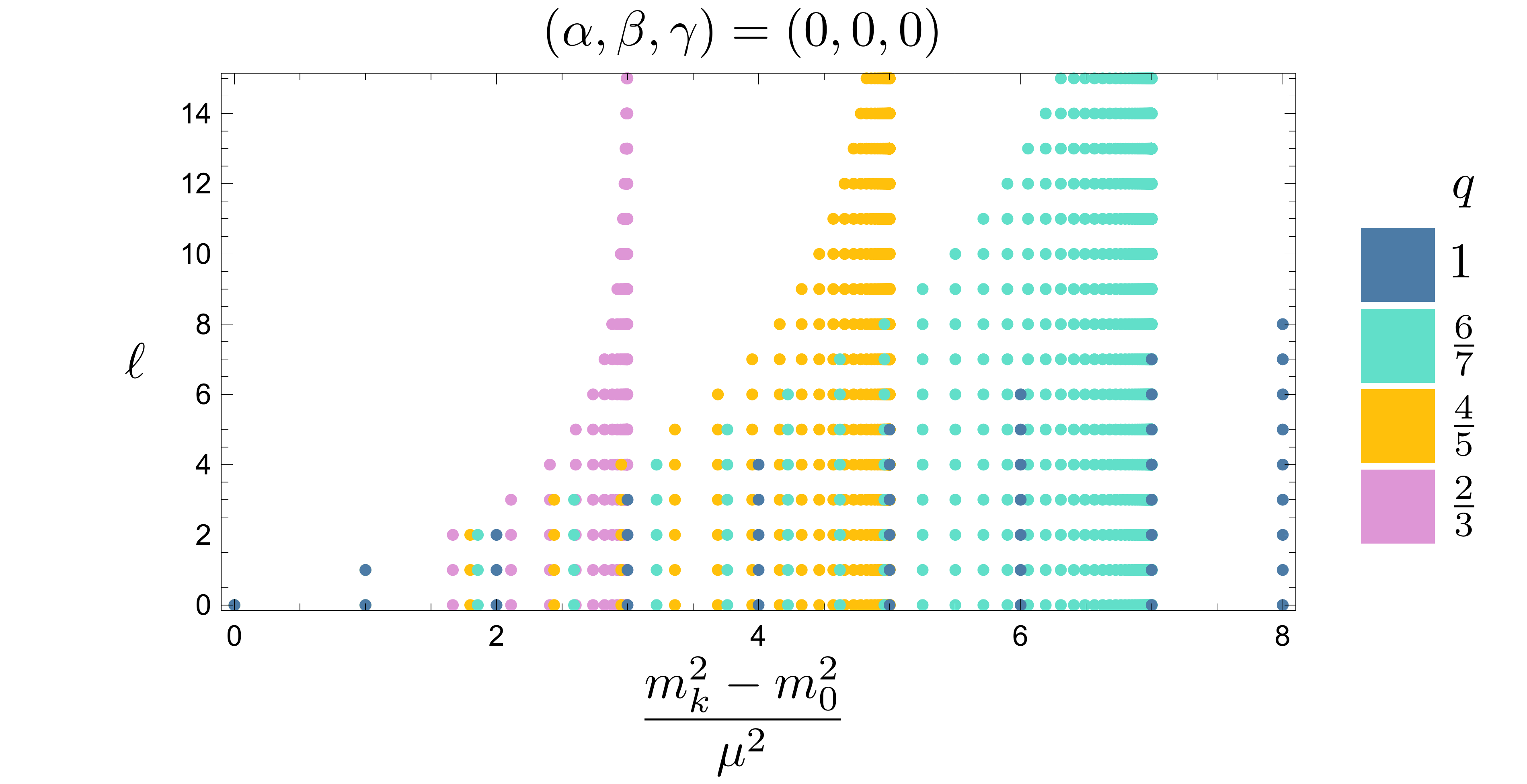}\\\vspace{5mm}
\includegraphics[width=11cm]{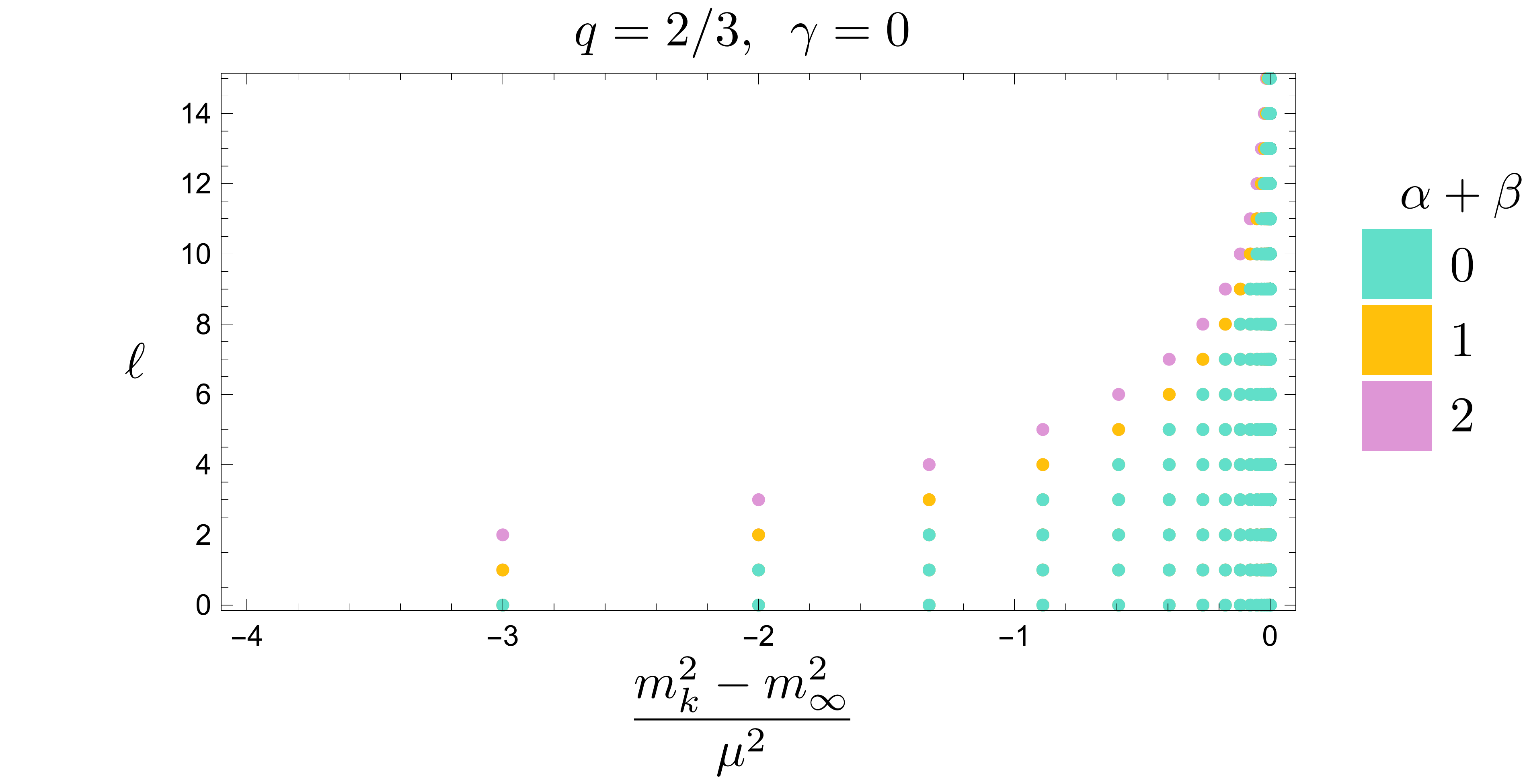}\\\vspace{5mm}
\includegraphics[width=11cm]{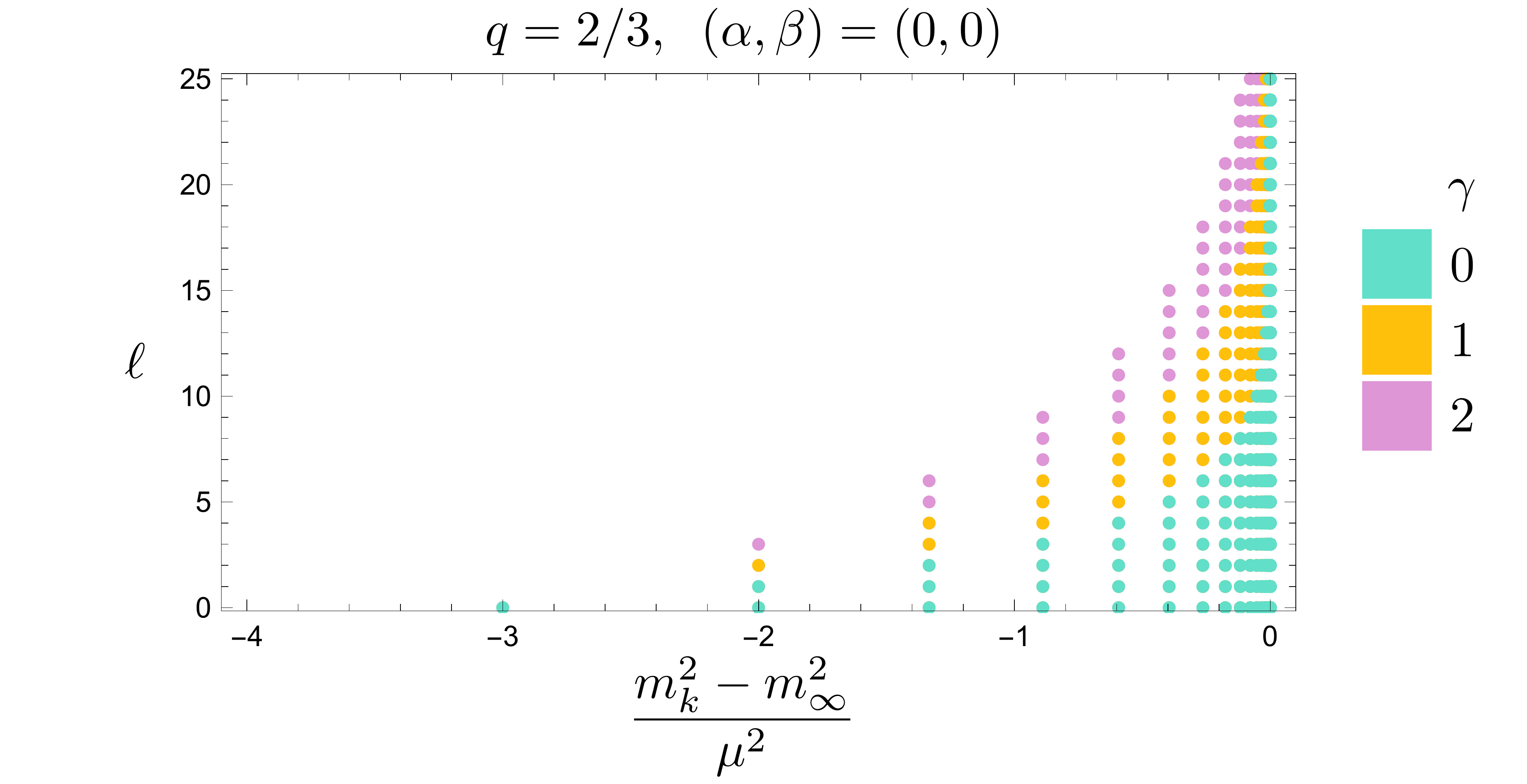}
\end{center}
\vspace{-0.5cm}
\caption{Chew-Frautschi plots depicting the spin $\ell$ and mass squared $m_k^2$ for the modes exchanged at the $k$th level in our amplitudes. {\it Top:} For $(\alpha,\beta,\gamma)=(0,0,0)$ and various values of $q$, we see how the logarithmic Regge trajectories and accumulation points of the original Coon amplitude ($q<1$) interpolate into the linear trajectories and unbounded spectrum of the Veneziano amplitude ($q=1$).
{\it Bottom left:} For $q=2/3$, different values of $\alpha+\beta$ add linearly to $\ell_{\max}(k)$, though $\alpha$ and $\beta$ enter independently in the actual amplitude.
{\it Bottom right:} For $q=2/3$, different values of $\gamma$ add multiplicatively to $\ell_{\max}(k)$. Note that the colored dots are overlayed, so it should be understood that the states represented by pink, yellow, and teal have spins ranging from $\ell_{\max}(k)$ down to 0 and all first appear at $m_0^2$. 
}
\label{fig:ChewFrautschi}
\end{figure}

On the other hand, in our amplitude, the pattern of exchanged spins differs markedly from that of string theory.  Recall that on the $k$th pole at $\sigma \rightarrow q^k$, the residue $R_k$ is a polynomial in $\tau$ with powers ranging from $\gamma k+\beta$ to $(1+\gamma)k+\alpha+\beta$.  After recasting $\tau$ in terms of $t$ using \Eq{eq:sigmatau}, we see that $R_k$ becomes a polynomial in $t$ with powers ranging from $0$ to\footnote{Note that for the special case of $\delta=1/(q-1)$, the spin ranges from $\gamma k + \beta$ to $\ell_{\max}(k)$. } 
\eq{
\ell_{\rm max}(k)=(1+\gamma)k+\alpha+\beta.
}{eq:lmax}
In summary, we have constructed a family of amplitudes in which the intercept and slope of the Chew-Frautschi plot for the $(\alpha,\beta,\gamma)$ theory---plotting the highest-spin state as a function of pole number $k$---is customizable.
As in the Coon amplitude, the Regge trajectory, in terms of maximum spin as a function of energy, is logarithmic by Eq.~\eqref{eq:spoles}, rather than the linear trajectory of Veneziano.  
See Fig.~\ref{fig:ChewFrautschi} for an illustration.

\subsubsection{Kinematic Limits}

Next, let us consider the amplitude in various kinematic limits.  We begin with the low-energy limit, which we take to mean the kinematic regime in which the external momenta are close to the mass of the lowest-lying modes in the spectrum.  This corresponds to taking $\sigma,\tau\rightarrow 1$ in \Eq{eq:Afull}, yielding the infrared amplitude,
\eq{
A_{\rm IR}(\sigma,\tau) = \lim_{\sigma,\tau\rightarrow 1}A(\sigma,\tau) = -c_{\alpha\beta\gamma}\frac{(q^{-\alpha};q)_\alpha}{(q;q)_\infty} \left(\frac{1}{1-\sigma}+\frac{1}{1-\tau}\right)+\cdots,
}{}
where we have defined the $q$-Pochhammer symbol,
\eq{
(r;q)_n &= \prod_{k=0}^{n-1}(1-rq^k) .
}{eq:qPoch}
In the case where the lowest-lying modes are massless, this corresponds to $A_{\rm IR}(s,t) \sim \tfrac{1}{s}+\tfrac{1}{t}$, which is the amplitude of massless cubic scalar theory.\footnote{Note that the would-be double pole in the expansion, scaling as $(1\,{-}\,\sigma)^{-1}(1\,{-}\,\tau)^{-1}$, has a coefficient $\sim c_{\alpha\beta\gamma} (q^{-\alpha};q)_\infty/(q;q)_\infty^2$, which vanishes for $\alpha\geq 0$. Equivalently, in the notation we will introduce in Sec.~\ref{sec:VenezianoRelation}, the would-be $1/st$ pole for massless external states has a coefficient $\sim d_{\alpha\beta\gamma}q^\alpha/\Gamma_q(-\alpha)$, which vanishes for nonnegative integers $\alpha$ due to the poles in the $q$-gamma function.}

Furthermore, we can trivially compute the behavior of the amplitude at high-energy, fixed-angle scattering, since the infinite product in \Eq{eq:Afull} trivially asymptotes to unity.  Taking $\sigma,\tau \rightarrow -\infty$ in \Eq{eq:Afull}, we obtain the following scaling for the ultraviolet amplitude,
\eq{
\log A_{\rm UV}(\sigma, \tau) &= \lim_{\sigma,\tau\rightarrow -\infty} \log A(\sigma,\tau)\\
& = \log \left[c_{\alpha\beta\gamma} (\sigma\tau)^{\alpha+\beta} q^{(1+\gamma)\frac{\log \sigma}{\log q}\frac{\log \tau}{\log q}} \right] + \cdots\\
&= (\alpha+\beta)(\log \sigma + \log \tau) + (1+\gamma) \frac{ \log \sigma \log \tau}{\log q} +\cdots,
}{eq:AUV}
which is of course nothing more than the logarithm of the $W$ prefactor. 
The result in \Eq{eq:AUV} implies that at intermediate high energies the amplitude has poor but still polynomial behavior, $A_{\rm UV}\sim (\sigma\tau)^{\alpha+\beta}$.  However, at extreme high energies, far above the accumulation point, the amplitude exhibits softened behavior, $A_{\rm UV} \sim \exp\left[(1+\gamma) \tfrac{\log\sigma \log\tau}{\log q}\right]$.  Note that since $q<1$  and $\gamma\geq 0$, the coefficient of $\log \sigma\log\tau$ is negative, so the ultraviolet behavior improves indefinitely.  By the same logic, the Regge growth is $A_{\rm Regge} \sim \sigma^{\alpha+ \beta+ (1+\gamma)\log \tau/\log q}$.  If we impose Regge boundedness, then $\alpha+\beta\leq 2$, which as we will see later is also apparently required by partial wave unitarity in conjunction with imposing additional physical requirements.

In our analysis we have assumed throughout that the external states are scalars.  But by multiplying these scalar amplitudes by polarization-dependent kinematic factors, we can easily derive putative  expressions for higher-spin external states.  That said, this procedure will invariably introduce additional powers of momenta, so the above infrared and ultraviolet expressions asymptotics will all be modified.  In particular, there is a general tension between lifting to higher-spin external states and maintaining Regge boundedness.

\subsubsection{Branch Cuts}

As we have seen in \Eq{eq:AUV}, our new amplitudes exhibit branch cuts ending at the accumulation point in the spectrum. As we noted earlier, Ref.~\cite{Caron-Huot:2016icg} closely examined the uniqueness of the Veneziano amplitude but specifically assumed the absence of accumulation points in their arguments.  That said, Ref.~\cite{Caron-Huot:2016icg} did raise concerns about unitarity of the Coon amplitude on account of its double branch cut at high energies of the form 
\be 
\log A_{\rm UV}(s,t) \sim -\log(-s) \log (-t),
\label{eq:highenergy}
\ee
which also appears in our amplitudes.
However, as calculated in Ref.~\cite{Maldacena:2022ckr} using the results of Ref.~\cite{Alday:2007hr}, string theory itself exhibits amplitudes with precisely this behavior, arising from the scattering of open strings ending on a D-brane in AdS.  
We therefore cannot rule out any of the accumulation-point generalizations of the Veneziano amplitude on the basis of branch cut singularities, provided string theory is consistent.
Instead, we check unitarity directly by computing the residues of all partial waves and demanding positivity, as in Sec.~\ref{sec:positivity}.

\subsubsection{Relation to Coon Amplitude}

The original analysis of Coon \cite{Coon} improperly omitted the amplitude prefactor $W(\sigma, \tau)$, but this was later rectified in Refs.~\cite{Coon:1972qz, Tourkine}.  All of these past results correspond to a special case of our amplitudes for $(\alpha,\beta,\gamma)=(0,0,0)$.  

Conversely, we can recast the amplitude for general $(\alpha,\beta,\gamma)$ as a natural multi-parameter deformation of the Coon amplitude.  In particular, defining
\be
\begin{aligned}
\sigma &= q^{-\alpha} \tilde\sigma  \\
\tau &= q^{-\alpha}\tilde\tau \\
q &= \tilde q^{1+\gamma},
\end{aligned}
\ee
then our amplitude in \Eq{eq:Afull} becomes
\be
A(\sigma,\tau) =  \tilde c_{\alpha\beta\gamma} (\tilde \sigma\tilde\tau)^{\beta-\alpha\gamma}\tilde q^{\frac{\log \tilde\sigma}{\log \tilde q}\frac{\log \tilde\tau}{\log \tilde q}} \prod_{n=0}^\infty \frac{\left(1-\frac{\tilde q^{(1+\gamma)(n+\alpha)}}{\tilde\sigma\tilde\tau}\right)}{\left(1-\frac{\tilde q^{(1+\gamma)(n+\alpha)}}{\tilde\sigma}\right)\left(1-\frac{\tilde q^{(1+\gamma)(n+\alpha)}}{\tilde\tau}\right)}.
\ee
This expression is literally the Coon amplitude expressed in terms of $\tilde \sigma$ and $\tilde \tau$, with two small differences: a) there is an additional factor of $ (\tilde \sigma\tilde\tau)^{\beta-\alpha\gamma}$ out front, and b) the infinite product effectively runs over $(1+\gamma)(\mathbb{N}_0 +\alpha)$ rather than $\mathbb{N}_0$.  

By construction, every residue of the Coon amplitude on a pole in $\sigma$ yields a polynomial in $\tau$. The deformation in a) clearly preserves this property, since it multiplies the Coon amplitude by an overall monomial in $\sigma$ and $\tau$.  Furthermore, we could have more directly arrived at the deformation in b) by solving the polynomial residue constraint for the  ansatz in \Eq{eq:prod} but with the product running over $(1+\gamma)\mathbb{N}_0 +\alpha$ rather than $\mathbb{N}_0$. Note that the same is {\it not true} for a product ansatz running over an arbitrary subset of the integers.  In this more general case the residues will be nonpolynomial.

\subsubsection{Relation to Veneziano Amplitude}\label{sec:VenezianoRelation}

In order to relate our amplitudes to those of Veneziano, let us examine specific values of $q$. To ensure smooth limits as $q\rightarrow 0$ and $q\rightarrow 1$, we rescale the overall normalization of the amplitude $c_{\alpha\beta\gamma}$ by
\be 
c_{\alpha\beta\gamma} = d_{\alpha\beta\gamma} (1-q)^{1-\alpha} q^\alpha \prod_{n=0}^\infty (1-q^{n+1}).
\ee
Next, recall the definition of the $q$-gamma function,
\eq{
\Gamma_q(x) &= \frac{(q;q)_\infty}{(q^x;q)_\infty} (1-q)^{1-x},
}{}
expressed in terms of the $q$-Pochhammer symbol in \Eq{eq:qPoch}.
The latter reduces to the usual gamma function $\Gamma(x)$ for $q\rightarrow 1$.  In terms of these functions, we obtain the simple expression
\eq{
A(\sigma, \tau) &= d_{\alpha\beta\gamma}q^{\alpha}(\sigma\tau)^{\alpha+\beta}q^{(1+\gamma)\frac{\log\sigma}{\log q}\frac{\log\tau}{\log q}}\frac{\Gamma_{q}\left(-\frac{\log\sigma}{\log q}\right)\Gamma_{q}\left(-\frac{\log\tau}{\log q}\right)}{\Gamma_{q}\left(-\alpha-\frac{\log\sigma}{\log q}-\frac{\log\tau}{\log q}\right)},
}{eq:Afull_alt}
which is just a rewriting of the amplitude in \Eq{eq:Afull}.

Now let us take the $q\rightarrow 1$ limit.   First, we note that the mass spectrum is obtained from \Eq{eq:spoles} by sending the $q$-deformed integers $[n]_q$ to the regular integers $n$, so
\be
m_n^2 = \mu^2 \left(\delta + n\right).
\ee
This is precisely the evenly spaced and unbounded mass spectrum of the Veneziano amplitude.
However, as we will soon see, the distribution of spins is markedly different.

Second, we take the $q\rightarrow 1$ limit of the amplitude in \Eq{eq:Afull_alt}, which is by construction smooth.  Since $\sigma,\tau\rightarrow 1$ when $q\rightarrow 1$, the dependence on $\beta$ from the prefactor disappears in the limit. Further, using that  $\Gamma_q(-\tfrac{\log\sigma}{\log q}) \overset{q\rightarrow 1}{=} \Gamma(-\tfrac{s}{\mu^2}+\delta)$, we find that
\be
A(s, t) \overset{q\rightarrow 1}{=} d_{\alpha\beta\gamma}\frac{\Gamma\left(-\frac{s}{\mu^{2}}+\delta\right)\Gamma\left(-\frac{t}{\mu^{2}}+\delta\right)}{\Gamma\left(-\alpha-\frac{s+t}{\mu^{2}}+2\delta\right)},\label{eq:almostVeneziano}
\ee
where we have also transformed to the physical $s$ and $t$  Mandelstam variables.  
For $\alpha=0$, the amplitude in \Eq{eq:almostVeneziano} is literally the Veneziano amplitude, 
where $1/\mu^2$ and $-\delta$ play the respective roles of $\alpha'$ and $\alpha_0$.  
Remarkably, for $\alpha\neq0$ we obtain a {\it new deformation} of the Veneziano amplitude that exhibits the usual string spectrum---that is, without accumulation points---but encodes a {\it shifted} distribution of spin exchanges.  In particular, at the $k$th mass level, we observe a tower of exchanged modes with spin $0$ to $k+\alpha$.

\section{Unitarity Bounds}  
\label{sec:positivity}

In this section we derive positivity bounds on our amplitudes following from partial wave unitarity.   We begin with an analytic approach, deriving positivity bounds that are simple to understand before constructing the general case.  Afterwards, we numerically evaluate the general positivity bounds in order to carve out the space of consistent theories.

\subsection{Analytic Positivity}

Our starting point is the residue of the four-parameter amplitude in \Eq{eq:Afull} evaluated at the simple pole at $\sigma \rightarrow q^k$, yielding the compact expression\footnote{Modulo positive constant factors, this formula extracts the residue on a physical factorization channel, so $ \lim\limits_{\sigma\rightarrow q^k}(1-\tfrac{q^k}{\sigma})A(\sigma,\tau) \sim \lim\limits_{s\rightarrow m_k^2} (m_k^2-s) A(s,t)$.}  
\eq{
R_k(\tau)&=
 \lim_{\sigma\rightarrow q^k}\left(1-\tfrac{q^k}{\sigma}\right)A(\sigma,\tau)
= c_{\alpha \beta\gamma} q^{k(\alpha+\beta) } \tau^{(1+\gamma)k+ \alpha+\beta } \frac{\prod\limits_{n=0}^{k+\alpha-1} (1-\tfrac{q^n}{\tau q^{k+\alpha} } )}{\prod_{j=0,j\neq k}^{\infty} (1-\tfrac{q^{j}}{q^k})}.
}{eq:Rkproduct}
As advertised, this residue is a polynomial in $\tau$.

To streamline our positivity analysis, we will simplify equations by using $\sim$ to denote expressions that have the same sign, modulo magnitudes, so for example 
\eq{
q^{k(\alpha+\beta)} \sim 1 \qquad \textrm{and} \qquad
\prod_{j=0,j\neq k}^{\infty} (1-\tfrac{q^{j}}{q^k})
\sim (-1)^k.
}{}
Dropping all factors that do not affect the overall sign, we obtain
\eq{
R_k(\tau)
&\sim c_{\alpha\beta\gamma} (-1)^k \tau^{(1+\gamma)k+ \alpha+\beta }  ( \tfrac{1}{\tau q^{k+\alpha}} ; q)_{k+\alpha} \\
&\sim c_{\alpha\beta\gamma} (-1)^k \tau^{(1+\gamma)k+ \alpha+\beta }  \sum\limits_{j=0}^{k+\alpha}  \genfrac[]{0pt}{1}{k+\alpha}{j}_q (-\tfrac{1}{\tau q^{k+\alpha}})^j q^{\genfrac[]{0pt}{1}{j}{2}}\\
&\sim c_{\alpha\beta\gamma}  \sum\limits_{j=0}^{k+\alpha} (-1)^{j+k}   \genfrac[]{0pt}{1}{k+\alpha}{j}_q   q^{ \genfrac[]{0pt}{1}{j}{2}-j(k+\alpha)}   \tau^{(1+\gamma)k+ \alpha+\beta-j}.
}{eq:Rk_gen}
In the last line we have expanded the residue to obtain the precise coefficients of each term in the $\tau$ expansion.  Here we have used the following formulas for the $q$-Pochhammer symbol in terms of $q$-binomial coefficients, 
\eq{
(r; q)_n &= \sum\limits_{j=0}^n \genfrac[]{0pt}{1}{n}{j}_q (-r)^j q^{\genfrac[]{0pt}{1}{j}{2}} \qquad \textrm{where} \qquad
\genfrac[]{0pt}{1}{n}{j}_q = \frac{[n]!_q}{[n-j]!_q [j]!_q}=\prod_{i=0}^{j-1} \frac{1-q^{n-i}}{1-q^{i+1}},
}{eq:qbinomial}
where the $q$-factorial is $[n]_q!=(q;q)_n/(1-q)^n$ and $\genfrac[]{0pt}{1}{n}{j}$ is the standard binomial coefficient.

In order to evaluate the positivity of each partial wave, we recast the $\tau$-dependence residue in terms of the physical scattering angle $\theta$ via 
\eq{
\cos\theta=x = 1 + \frac{2t}{s-4\mu^2 \delta}.
}{eq:costheta}  
As previously noted, following the case of the Veneziano amplitude and the unitarity analysis of Ref.~\cite{Tourkine}, we have {\it assumed} that the masses of the external states are equal to those of the lowest-lying state in the exchanged tower, namely, $m_0^2=\mu^2\delta$.   
With this relation, we rewrite the auxiliary kinematic variable on the pole $\sigma\rightarrow q^k$, defining parameters akin to the Coon amplitude analysis of Ref.~\cite{Tourkine},
\be 
\begin{aligned}
\tau &= -b_k (x-x_{\infty,k}) \\
b_{k} & =\frac{1-q^{k}+3\delta(q-1)}{2}\\
x_{n,k} & =\frac{3-q^{k}+\delta(q-1)-2q^{n-k-\alpha}}{1-q^{k}+3\delta(q-1)},
\end{aligned}\label{eq:AB}
\ee 
where we define $x_{n,k}$ for later convenience and $x_{\infty,k}=\lim_{n\rightarrow \infty}x_{n,k}$. Written in terms of $x$ defined in Eq.~\eqref{eq:costheta}, the residue has dependence on five parameters, $(q, \alpha,\beta,\gamma,\delta)$.  The sixth parameter $\mu^2$, defined in \Eq{eq:sigmatau}, drops out of the partial wave unitarity analysis as it just sets the overall scale of all Mandelstam invariants.

Next, we define the partial wave decomposition as
\eq{
R_k(x) = \sum_\ell a_{k,\ell} G_\ell^{(D)} (x),
}{eq:Rkpartials}
where the Gegenbauer polynomial\footnote{In conventional mathematical notation, our function $G_\ell^{(D)}(x)$ is written as $C_\ell^{(\frac{D-3}{2})}(x)$.} $G_\ell^{(D)}$ corresponds to a spin-$\ell$ partial wave in $D$ spacetime dimensions.  The coefficient of each partial wave is
\be
a_{k,\ell} =\frac{2^{D-4}\left(\frac{D-3}{2}+\ell\right)\ell![\Gamma\left(\frac{D-3}{2}\right)]^{2}}{\pi\Gamma(\ell+D-3)}\int_{-1}^{+1} {\rm d}x \,  (1-x^2)^{\frac{D-4}{2}}  G_\ell^{(D)} (x) R_k(x),
\ee
and a necessary and sufficient condition for partial wave unitarity is that 
\eq{
a_{k,\ell} \geq 0,
}{}
so the spectrum does not contain negative-norm states.

\subsubsection{Simple Bounds}\label{sec:specific}

Before conducting a general analysis of all partial waves, we first derive a few simple bounds that can be extracted straightforwardly from the above formulas.  To be concrete, we compute the leading and subleading partial wave coefficients, $a_{k,\ell_{\rm max}(k)}$ and $a_{k,\ell_{\rm max}(k)-1}$, derived from the residue on the $k$th level. Partial wave unitarity enforces that these coefficients are positive, thus carving out a restricted region of putatively consistent theories in the $(q,\alpha,\beta,\gamma, \delta)$ parameter space.

Expanding the residue in \Eq{eq:Rk_gen} to its leading and subleading powers of $\tau$, we obtain 
\eq{
R_k(\tau)
\sim c_{\alpha\beta\gamma} (-1)^{k}  \tau^{\ell_{\rm max}(k)}\left(1- \frac{[k+\alpha]_q}{q^{k+\alpha}\tau}+\cdots\right).
}{eq:Rktauexpansion}
Substituting the $\tau$-dependence for $x$ via \Eq{eq:AB} and focusing on the leading partial wave coefficient, we obtain
\eq{
R_k(x) &\sim c_{\alpha\beta\gamma}(-1)^k (  3\delta-[k]_q)^{\ell_{\rm max}(k)} x^{\ell_{\rm max}(k)} + \cdots.
}{}
Using that the coefficient of $x^{\ell_{\rm max}(k)}$ is proportional to $a_{k,\ell_{\rm max}(k)}$, we find
\eq{
a_{k,\ell_{\rm max}(k)} &\sim c_{\alpha\beta\gamma}(-1)^k ( 3\delta-[k]_q)^{\ell_{\rm max}(k)}\\
&\sim c_{\alpha\beta\gamma}(-1)^{k} (3\delta(1-q) - 1+ q^k)^{(1+\gamma)k+ \alpha+\beta}.
}{}
Since there must be $k$ such that  $q^{k}-1+3\delta(1-q)\neq 0$, it follows that if $\gamma$ is odd, there exist $k$ for which $[q^k-1+3\delta(1-q)]^{(1+\gamma)k}>0$, which in turn would imply that $a_{k,\ell_{\max}(k)}\sim (-1)^k$ times a $k$-independent constant at large $k$.  This is inconsistent with positivity, so $\gamma$ must be even.

Unitarity also forbids $\delta=\frac{1}{3(1-q)}$, since such an equality would set $a_{k,\ell_{\max}(k)} \sim (-1)^k c_{\alpha\beta\gamma}$.
At very large $k$, where $q^k\rightarrow 0$, we have
\be
a_{k,\ell_{\max}(k)} \sim (-1)^{\alpha+\beta} c_{\alpha\beta\gamma}(1-3\delta(1-q))^{k+\alpha+\beta},
\ee
so we must have $c_{\alpha\beta\gamma}\sim(-1)^{\alpha+\beta}$ and $\delta< \frac{1}{3(1-q)}$.
Next, considering the case of general $k$, we find $a_{k,\ell_{\max}(k)} \sim ([k]_q -3\delta)^{k+\alpha+\beta}$, so we must have
\be
\delta \leq  \frac{1}{3}[k]_q\qquad \text{for all}\qquad \begin{cases} k \geq 1,&\alpha\;{=}\;\beta\,{=}\;0 \\ k\geq 0,&\alpha + \beta>0.
\end{cases}
\ee
In summary, we conclude that
\be
\text{$\gamma$ is even,}\qquad c_{\alpha\beta\gamma} \sim(-1)^{\alpha+\beta},\qquad\textrm{and} \qquad\delta \leq \begin{cases} \frac{1}{3},&\alpha\;{=}\;\beta\,{=}\;0 \\ 0,&\alpha+\beta >0, \end{cases}\label{eq:necessaryleading}
\ee
which are the necessary conditions implied by unitarity of the leading partial wave coefficient.

Similarly, the coefficient of $x^{\ell_{\max}(k)-1}$ in the expansion of $R_k(x)$ gives the subleading partial wave coefficient $a_{k,\ell_{\max}(k)-1}$ up to a postive factor.
From the $\tau$ expansion in Eq.~\eqref{eq:Rktauexpansion} and the binomial expansion of $\tau^{\ell_{\max}(k)-1}$ in terms of $x$ using Eq.~\eqref{eq:AB}, we have
\be
a_{k,\ell_{\max}(k)-1} \sim c_{\alpha\beta\gamma}(-1)^{k+\ell_{\max}(k)} (b_k)^{\ell_{\max}(k)}\left(-\ell_{\max}(k) x_{\infty,k} + \frac{[k+\alpha]_q}{ b_k q^{k+\alpha}}\right).
\ee
Using the results of Eq.~\eqref{eq:necessaryleading}, $c_{\alpha\beta\gamma}(-1)^{k+\ell_{\max}(k)} \sim +1$ and $b_k\sim +1$, so
\be
a_{k,\ell_{\max}(k)-1} \sim  2[k+\alpha]_q -q^{k+\alpha}\ell_{\max}(k)\left(3-q^{k}+\delta(q-1)\right).
\ee
Positivity therefore requires
\be 
\delta \geq \frac{2}{1-q} + [k]_q -\frac{2[k+\alpha]_q}{(1-q)q^{k+\alpha} \ell_{\max}(k)}.\label{eq:subleading}
\ee
Among $k\geq 1$, the strictest bounds from Eq.~\eqref{eq:subleading} appear to always come from $k=1$. We have the necessary condition
\be
\delta \geq 1+ \frac{2}{1-q} - \frac{2[1+\alpha]_q}{(1+\alpha+\beta+\gamma)q^{1+\alpha}(1-q)},
\ee
which, given the upper bound on $\delta$ in Eq.~\eqref{eq:necessaryleading}, implies a maximum value of $\beta+\gamma$ as a function of $q$ and $\alpha$.

Deriving unitarity bounds for ever-more subleading partial waves entails expanding \Eq{eq:Rk_gen} to higher and higher orders, which quickly becomes intractable.
We will therefore turn to a more general analysis in the subsequent sections, computing the explicit form of all partial wave coefficients analytically and then exploring this space numerically.

Before moving on, let us derive a set of {\it sufficient} conditions guaranteeing partial wave unitarity for $\beta=\gamma=0$ and arbitrary $\alpha$.
In particular, we observe from \Eq{eq:Rkproduct} that for this choice of parameters, $R_k \sim (-1)^\alpha c_{\alpha00}\prod_{n=0}^{k+\alpha-1} (-\tau+ q^{n-k-\alpha})$, with no extra factors of $\tau$.
Thus a sufficient condition to ensure positivity of all partial waves is to require that $R_k$ be a polynomial in $x$ with all positive coefficients.
Using that $c_{\alpha 0 0} \sim (-1)^\alpha$ from Eq.~\eqref{eq:necessaryleading} and the parameterization in Eq.~\eqref{eq:AB}, we have $R_k \sim \prod_{n=0}^{k+\alpha-1} b_k(x-x_{n,k})$, so we can enforce positivity by demanding that $b_k\geq 0$ and $x_{n,k}\leq 0$ for all $n=0,\ldots,k+\alpha-1$.
In terms of the parameters defining our amplitude, these conditions imply
\be
[k]_q + 2[n-k-\alpha]_q \leq \delta \leq \frac{1}{3}[k]_q, 
\ee
which corresponds to the following sufficient condition for partial wave unitarity for all theories with $\beta=\gamma=0$ and arbitrary $\alpha$:
\be
c_{\alpha 00} \sim (-1)^{\alpha} \;\;\;\text{and}\;\;\; -\frac{2}{q} + \frac{1}{1-q} \leq \delta \leq 0.
\ee
For the special case of $\alpha=0$, the lower bound for $\delta$ given above matches the sufficient positivity condition derived for the Coon amplitude in Ref.~\cite{Tourkine}.

\subsubsection{General Bounds}

For the sake of completeness, let us now derive a closed formula for all partial wave coefficients of our amplitudes on their residues.  To accomplish this task it will be convenient to introduce a compact generating function of all Gegenbauer polynomials,
\eq{
(1-2x \lambda + \lambda^2)^{-\frac{D-3}{2}} &= \sum_{\ell=0}^\infty \lambda^\ell G_\ell^{(D)} (x).
}{}
Convolving this generating function with the residue $R_k(x)$, we obtain a generating function for all partial wave coefficients,
\eq{
 R_k(\lambda) &= \int_{-1}^{+1} {\rm d}x \, R_k(x) (1-2x \lambda + \lambda^2)^{-\frac{D-3}{2}}\\
&\sim c_{\alpha\beta\gamma} (-1)^{\ell_{\max}(k)-k}   \sum\limits_{j=0}^{k+\alpha}   \genfrac[]{0pt}{1}{k+\alpha}{j}_q   q^{ \genfrac[]{0pt}{1}{j}{2}{-}j(k{+}\alpha)} \! \sum_{i=0}^{\ell_{\max}(k)-j}\!\genfrac[]{0pt}{1}{\ell_{\max}(k)-j}{i} Q_i(\lambda)\tfrac{ (-b_k x_{\infty,k})^{\ell_{\max}(k)-j}}{(-x_{\infty,k} )^{i}} ,
}{eq:Rkgenerating}
\parbox{\textwidth}{in terms of $\ell_{\max}(k)$ in Eq.~\eqref{eq:lmax} and $b_k$, $x_{\infty,k}$ in \Eq{eq:AB}, and where we have defined the \parfillskip=0pt} \pagebreak

\noindent function $Q_i(\lambda)$ as
\eq{
 Q_i(\lambda)& =\int_{-1}^{+1} {\rm d}x \, (1-x^2)^{\frac{D-4}{2}} x^i (1-2x \lambda  + \lambda^2)^{-\frac{D-3}{2}}\\
&= \frac{\sqrt{\pi}}{2(1+\lambda^2)^{\frac{D-1}{2}}}\Biggl\{\tfrac{[1+(-1)^i](1+\lambda^2)}{2^i}\tfrac{\Gamma(1+i)\Gamma\left(\frac{D-2}{2}\right)}{\Gamma\left(1+\frac{i}{2}\right)\Gamma\left(\frac{D+i-1}{2}\right)} {}_3 F_2\left[\begin{array}{c} \frac{D-3}{4},\frac{D-1}{4},\frac{1+i}{2} \\ \frac{1}{2},\frac{D+i-1}{2} \end{array}; \tfrac{4\lambda^2}{(1+\lambda^2)^2}\right] \\
&\qquad\qquad\qquad\;\;\; + \tfrac{[1-(-1)^i]\lambda}{2^{D-4}} \tfrac{\Gamma\left(1+\frac{i}{2}\right)\Gamma(D-2)}{\Gamma\left(\frac{D-3}{2}\right)\Gamma\left(\frac{D+i}{2}\right)}{}_3 F_2\left[\begin{array}{c} \frac{D-1}{4},\frac{D+1}{4},1+\frac{i}{2} \\ \frac{3}{2},\frac{D+i}{2} \end{array}; \tfrac{4\lambda^2}{(1+\lambda^2)^2}\right]\Biggr\} 
\\
 &= \frac{\pi}{2^{D-4+i}}\frac{ \Gamma(D-3+i)}{\Gamma\left(\tfrac{D-3}{2}\right)\Gamma\left(\tfrac{D-1}{2} +i\right)} \times \lambda^i +\cdots,
}{}
where in the last line we have written the highest power in $\lambda$.
By unitarity, each partial wave must be positive, so 
\eq{
a_{k,\ell} \sim \lim_{\lambda\rightarrow 0}\frac{{\rm d}^\ell R_k(\lambda)}{{\rm d}\lambda^\ell} \geq 0.
}{}
Hence, by demanding positivity of all derivatives of $R_k(\lambda)$ at $\lambda=0$, we can carve out the space of unitary amplitudes.

Even more explicitly, by employing various identities involving the Gegenbauer polynomials and traces of nested sums, we can derive a closed-form expression for each partial wave coefficient, as shown in App.~\ref{app:partials}:
\be 
\begin{aligned}
a_{k,\ell} & =\left(1+\tfrac{2\ell}{D-3}\right)\Gamma\left(\tfrac{D-1}{2}\right)\tfrac{(-1)^{\ell}c_{\alpha\beta\gamma} q^{k(\alpha+\beta)}}{\left(q;q\right)_{\infty}\left(q^{-k};q\right)_{k}}\times \\&
\qquad\times \left(\sum_{s{=}0}^{\left\lfloor\!\frac{\beta{+}\gamma k{-}\ell}{2}\!\right\rfloor}\sum_{j{=}0}^{k{+}\alpha}+\!\!\!\!\sum_{s{=}\max\left\{\!0,1{+}\left\lfloor \!\frac{\beta{+}\gamma k{-}\ell}{2}\!\right\rfloor \!\right\}}^{\left\lfloor \!\frac{\alpha{+}\beta{+}(1{+}\gamma)k{-}\ell}{2}\!\right\rfloor}\sum_{j{=}\ell{+}2s{-}\beta{-}\gamma k}^{k{+}\alpha}\right)\\
 & \qquad\qquad\qquad \left[\tfrac{(j+\beta+\gamma k)![1-q^{k}+3\delta(q-1)]^{\ell+2s}[3-q^{k}+\delta(q-1)]^{j+\beta+\gamma k-\ell-2s}}{2^{\ell{+}2s{+}j{+}\beta{+}\gamma k}(j+\beta+\gamma k-\ell-2s)!s!\Gamma\left(\frac{D-1}{2}+\ell+s\right)}\tfrac{(q^{-k-\alpha};q)_{k{+}\alpha{-}j}}{(q;q)_{k{+}\alpha{-}j}}\right].
\end{aligned}\label{eq:aa}
\ee
The alternating signs in the sum mean that---as is notoriously the case for the Veneziano amplitude~\cite{Arkani-Hamed:2022gsa}---the partial waves are not {\it manifestly} positive. 
Reading off the unitary regions of parameter space will therefore require numerical analysis.

\subsection{Numerical Positivity}\label{sec:num}

At last, we are equipped to numerically ascertain the parameter space of our amplitudes that is consistent with partial wave unitarity.  To achieve this result, we simply evaluate $a_{k,\ell}$ using \Eq{eq:aa} for various choices of  $(q,\alpha,\beta,\gamma,\delta)$ and demand its positivity.  The space of parameters is vast, so to visualize the bounds we scan over various choices of the integers $(\alpha,\beta,\gamma)$ and plot the positive regions in the $(q,\delta)$ plane. 

\begin{figure}[!t]
\begin{center}
\includegraphics[width=\textwidth]{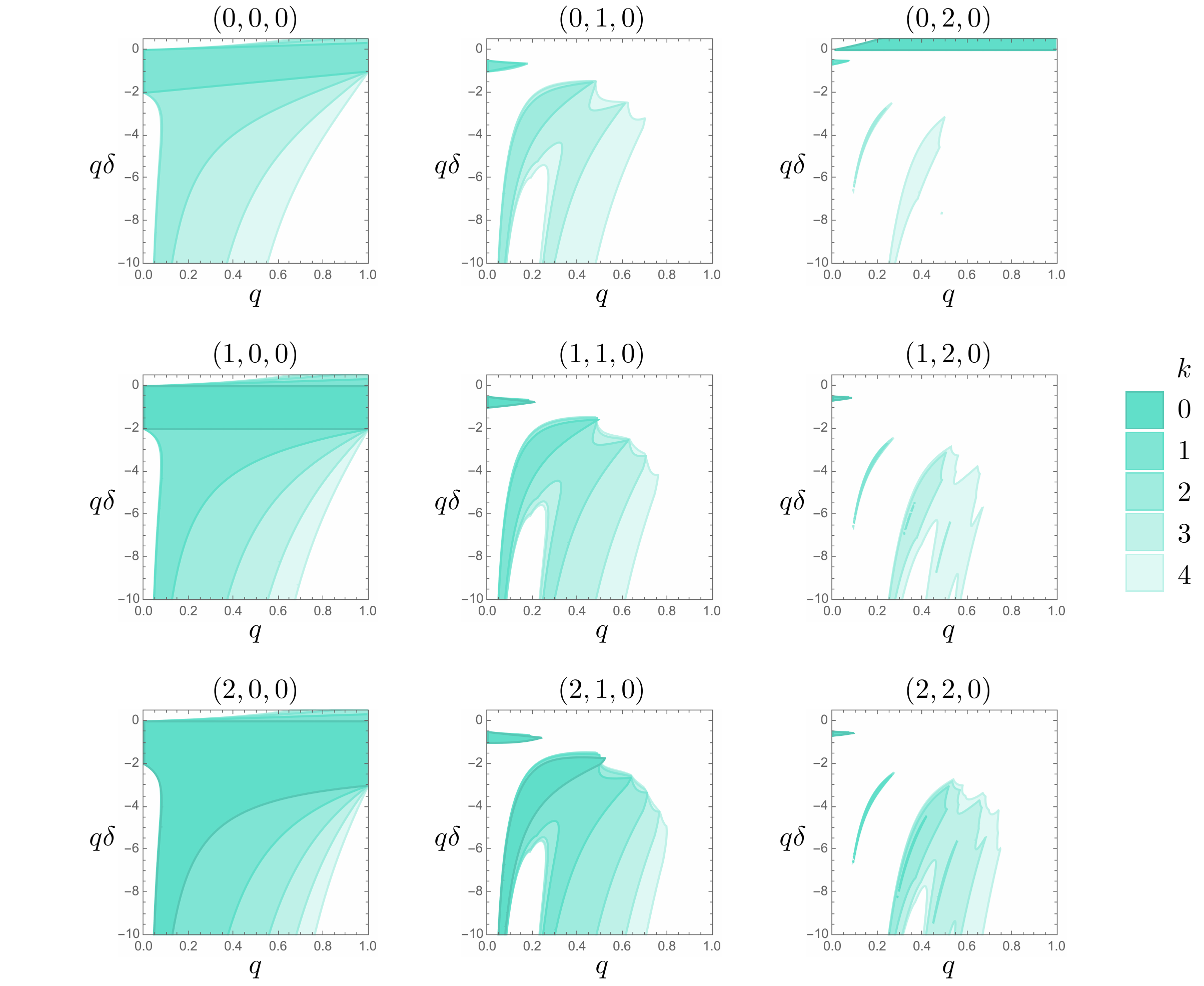}
\end{center}
\vspace{-0.5cm}
\caption{
Numerical plots depicting the parameter space for which all partial waves on the $k$th residue are unitary.  Positive regions are represented in the $(q, q\delta)$ plane for various choices of $(\alpha,\beta,\gamma)$ in $D=4$. Note that for the $(\alpha,\beta,\gamma)=(0,0,0)$ and $(0,1,0)$ theories, the $k=0$ pole is not depicted: for $(0,0,0)$ it is trivially positive, while for the $(0,1,0)$ case there exists no choice of $(q,\delta)$  consistent with positivity for $k=0$.
}
\label{fig:Boxes}
\end{figure}

In \Fig{fig:Boxes}, we have numerically evaluated the partial wave coefficients at the $k$th mass level for theories described by discrete choices of $\alpha,\beta \in \{0,1,2\}$ with $\gamma=0$, scanning over continuous values of $(q,\delta)$ and specializing to $D=4$ dimensions for concreteness.  We then plot the regions for which all the partial wave coefficients---of spin $\ell$ with $0\leq \ell\leq \ell_{\max}(k)$---at a given $k$th level are positive.  For ease of visualization, the plots in \Fig{fig:Boxes} are displayed in the $(q, q \delta)$ plane.
By numerical evaluation, we have not found any consistent regions with $\gamma \neq 0$.
  It would be interesting to see to what extent these amplitudes satisfy low-spin dominance~\cite{Bern:2021ppb}.
  
\begin{figure}[t]
\begin{center}
\includegraphics[width=\textwidth]{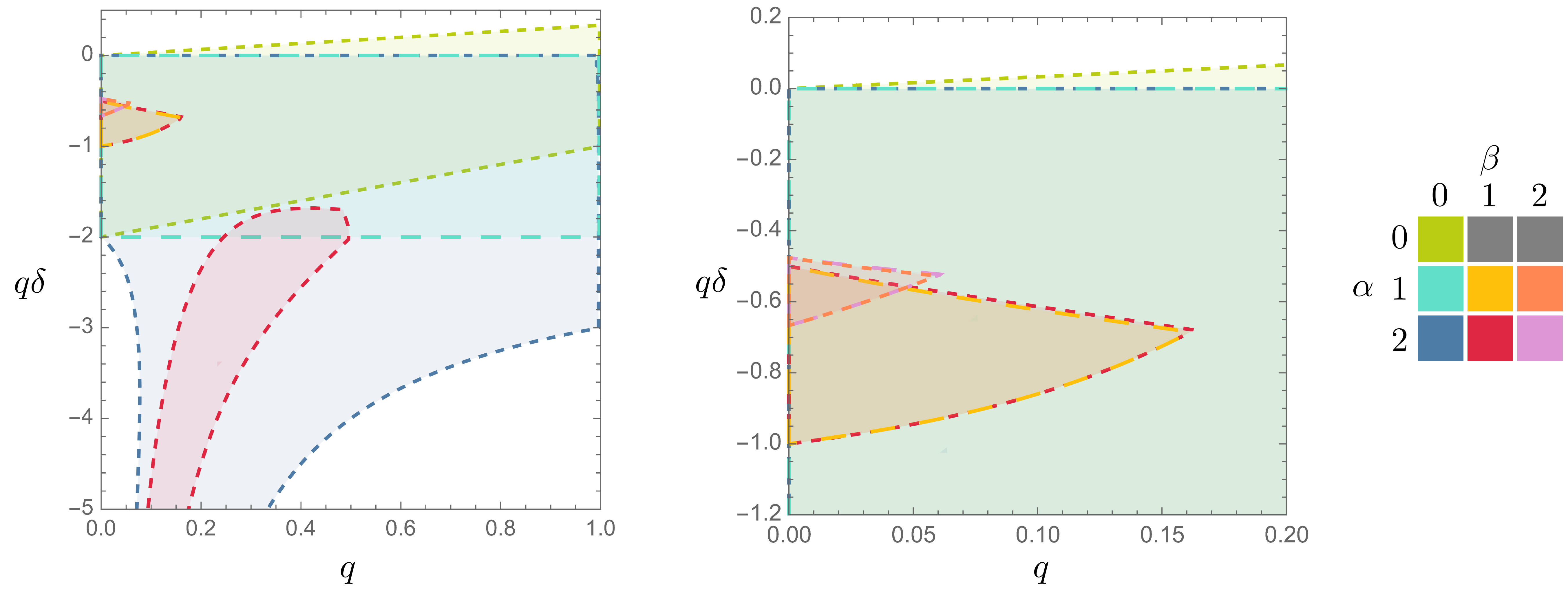}
\end{center}
\vspace{-0.5cm}
\caption{Numerical plots depicting the parameter space for which all partial waves on all calculated residues are unitary.  Positive regions are represented in the $(q, q\delta)$ plane for various choices of $\alpha,\beta \in \{ 0,1,2\}$, all with $\gamma=0$ and $D=4$. The cases $(\alpha,\beta)=(0,1)$ and $(0,2)$ evidently do not support any consistent region.\vspace{7mm}}
\label{fig:Regions}
\end{figure}

Meanwhile, \Fig{fig:Regions} depicts the {\it intersection} of all positivity bounds in \Fig{fig:Boxes}.  In particular, the colored area depicts the region in which all positivity bounds computed for $k\in\{0,\ldots,6\}$ are satisfied.  Higher values of $k$ do not appreciably affect this region.
Multiple consistent islands are present. 

Another physically important quantity is the critical spacetime dimension $D_{\rm crit}$, which defines the maximal dimension in which the amplitude remains unitary.  It is straightforward to compute $D_{\rm crit}$ for various choices of parameters using \Eq{eq:aa}.  For example, for the Veneziano amplitude, corresponding to $(q,\delta)=(1,0)$ and $(\alpha,\beta,\gamma)=(0,0,0)$, the critical dimension is $D_{\rm crit}=10$.  Interestingly, this quantity slowly rises with $\alpha$, so by incrementing up to $\alpha=2$ we obtain $D_{\rm crit}=11$, while going all the way to $\alpha=10$ gives $D_{\rm crit}=12$.  We leave a more complete investigation of these unitary islands to future work.

\subsection{Additional Spectral Constraints}

In principle there exist additional constraints on the space of allowed amplitudes based on the masses and spins appearing in the spectrum.  For example, we see from \Eq{eq:spoles} that for $\delta<0$, the lowest-lying mode in the spectrum is tachyonic.   Depending on the spins of these states, the presence of tachyonic instabilities can range from a mere nuisance to something more pathological.

For example, scalar tachyons famously arise in everything from the standard model to open string theory.  In such cases the scalar tachyons condense and there is no fundamental problem because we have improperly expanded about a saddle point or maximum.  While it is questionable whether one should even be computing amplitudes in such a background, this is certainly common practice when studying perturbative string tachyon amplitudes.

If, on the other hand, the tachyonic modes carry spin, then the corresponding instability results in ghost condensation~\cite{Arkani-Hamed:2003pdi} and the spontaneous breaking of Lorentz invariance~\cite{Thaler}.  Such theories exhibit extreme peculiarities, including violations of the second law of thermodynamics~\cite{Dubovsky:2006vk}. In principle, it might be logically possible to tame these issues if the ghost condensate points in the direction of an extra dimension.  

That said, a maximally conservative option is to simply forbid altogether any tachyons with spin, which we now assume.  On the $k=0$ pole, the amplitude exhibits the exchange of states of mass $\mu^2\delta$ and spin ranging from $0$ to $\alpha+\beta$.  Hence, the absence of spinning tachyons implies that if $\alpha+\beta>0$, then $\delta\geq 0$.  Meanwhile, the $k=1$ pole supports states of mass $\mu^2(\delta+1)$ and spins ranging $0$ to $\alpha+\beta+\gamma+1$, so we deduce that $\delta \geq -1$.
Combined with our positivity results in Eq.~\eqref{eq:necessaryleading}, we thus learn that $\delta \in [-1,\tfrac13]$ for $(\alpha,\beta)=(0,0)$, and $\delta=0$ otherwise.

Last but not least, we can also impose Weinberg's theorem~\cite{Weinberg:1964ew} forbidding the existence of interacting massless higher-spin particles with $\ell>2$.  Such states do not have a well-defined stress-energy tensor and cannot be consistently coupled to gravitons \cite{Huang:2013vha}.  The absence of states that are massless higher-spin or spinning tachyons then implies the bound $\alpha+\beta \leq 2$.

In summary, and incorporating our $D=4$ numerical results, we arrive at the following parameter space satisfying partial wave unitarity and free of spinning tachyons or massless higher-spin states:
\be 
\begin{aligned}
\delta \in [-1,\tfrac{1}{3}] &\qquad \textrm{for}\qquad (\alpha,\beta,\gamma)= (0,0,0)\\
\delta =0 \quad\;\; &\qquad \textrm{for}\qquad (\alpha,\beta,\gamma)= (1,0,0)\;\;{\rm or}\;\; (2,0,0),
\end{aligned}
\ee
where $q$ ranges freely from $0$ to $1$.
The first line defines the original Coon amplitude and the second encompasses representatives of our generalizations.  

\vspace{3mm}

\section{Unbounded Variations}
\label{sec:variations}

Up until now, our analysis has focused entirely on amplitudes exhibiting spectra with accumulation points at finite energy.  In this section we consider the opposite regime, in which the spectrum grows without bound.  We consider the cases of $st$-symmetric and $stu$-symmetric amplitudes, which should be interpreted as deformations of the cyclic and permutation invariant amplitudes of the open and closed string, respectively.

\subsection{Cyclic Invariant}

As our starting point, we consider the $st$-symmetric ansatz,
\be
A(\sigma,\tau)=W(\sigma,\tau)\prod_{n=0}^{\infty}\frac{1-\frac{\sigma+\tau}{g(n)}-\frac{\sigma\tau}{h(n)}}{\left(1-\frac{\sigma}{f(n)}\right)\left(1-\frac{\tau}{f(n)}\right)},\label{eq:proddiv}
\ee
which is reminiscent of \Eq{eq:prod} except that here the infinite products are convergent because the as yet undetermined functions $f(n)$, $g(n)$, and $h(n)$ are assumed to be {\it divergent} as $n\rightarrow \infty$.   As before, we take $f(n)$ to be monotonic without loss of generality.  Here the variables $\sigma$ and $\tau$ are related to the physical Mandelstam invariants $s$ and $t$ by an affine transformation.

Peeling off the first $p(k)$ factors of the numerator in analogy with Eq.~\eqref{eq:shift} and mandating cancellation of the remaining $\tau$-dependence in the product, we obtain a polynomial residue constraint that is identical to \Eq{eq:STC} except with $f$, $g$, and $h$ sent to their multiplicative inverses.  Thus, we again have $p(k)=k+\alpha$, for $\alpha$ a nonnegative integer, and the general solution is given by \Eq{eq:fghsolution_temp} except with $f\rightarrow 1/f$, $g\rightarrow 1/g$, and $h\rightarrow 1/h$.

Exploiting the freedom in affine transformations of $\sigma$ and $\tau$, we fix $f(0)=1$ and $f(1)=q$, where now $q>1$.  The resulting amplitude is
\be
A(\sigma,\tau)=W(\sigma,\tau)\prod_{n=0}^{\infty}\frac{1-\frac{\sigma\tau}{q^{n-\alpha}}}{\left(1-\frac{\sigma}{q^{n}}\right)\left(1-\frac{\tau}{q^{n}}\right)}. 
\ee
Unlike before, residues of the amplitude are local in $\tau$, so there is {\it not} a factor of $1/\tau^{k+\alpha}$ that needs to be cancelled by $W$, which we assume for simplicity to be monomial.  As a result,  we merely require that $W$ be $st$-symmetric and analytic in $\tau$ on the residue. This uniquely fixes the prefactor function $W(\sigma,\tau) = c_{\alpha\beta\gamma} (\sigma\tau)^{\beta}q^{\gamma\frac{\log\sigma}{\log q}\frac{\log\tau}{\log q}}$, where $\beta,\gamma$ are nonnegative integers.  The final form of the amplitude for a divergent spectrum is then
\be 
A(\sigma,\tau)=c_{\alpha\beta\gamma} (\sigma\tau)^{\beta}q^{\gamma\frac{\log\sigma}{\log q}\frac{\log\tau}{\log q}} \prod_{n=0}^{\infty}\frac{1-\frac{\sigma\tau}{q^{n-\alpha}}}{\left(1-\frac{\sigma}{q^{n}}\right)\left(1-\frac{\tau}{q^{n}}\right)}. \label{eq:Afulldiv}
\ee
Sending $q\rightarrow 1/q$, $\sigma\rightarrow 1/\sigma$, $\tau\rightarrow 1/\tau$, $\gamma\rightarrow -1-\gamma$, and $\beta\rightarrow -\alpha-\beta$, we arrive back to the accumulation-point amplitude in \Eq{eq:Afull}.

A preliminary analysis of partial wave unitarity for the amplitude in Eq.~\eqref{eq:Afulldiv} suggests that positivity requires $q\rightarrow 1$, corresponding to Eq.~\eqref{eq:almostVeneziano} and in agreement with previous analyses of the original Coon amplitude, which is nonunitary for $q>1$.  That is, it appears that the Veneziano amplitude of string theory and its generalizations in \Eq{eq:almostVeneziano} reside at the boundary of the consistent region, and constitute the only infinite-product amplitudes with an unbounded spectrum with polynomial residues.  
We leave a full analysis of this amplitude to future work.

\subsection{Permutation Invariant}  

Next, let us consider an analogous construction for a fully $stu$-symmetric amplitude.  In this case,  our underlying motivation is of course to investigate the uniqueness of the Virasoro-Shapiro amplitude of the closed string,
\be
M_{\rm string}(s,t) = -\frac{1}{stu}
\frac{\Gamma(-s)\Gamma(-t)\Gamma(-u )}{\Gamma(s)\Gamma(t)\Gamma(u)},\label{eq:VS}
\ee
where $s+t+u=0$ for massless external states and we have set all coupling constants and mass scales to unity and stripped off any external kinematic data associated with polarizations. 
Any sensible generalization of the Virasoro-Shapiro amplitude would be of particular interest, as it might offer an alternative path to unitarizing four-point graviton scattering.\footnote{While unitary sums of Virasoro-Shapiro amplitudes were constructed in Ref.~\cite{Arkani-Hamed:2020blm}, these are not of product form so we will not encounter them here.}

To this end, it is natural to generalize the approach of our earlier sections and consider a fully permutation invariant, infinite-product ansatz,
\be 
M(\sigma,\tau) = W(\sigma,\tau)\prod_{n=0}^{\infty}\frac{1-\frac{\sigma^2+\tau^2+\upsilon^2}{g(n)}-\frac{\sigma \tau \upsilon}{h(n)}}{\left(1-\frac{\sigma}{f(n)}\right)\left(1-\frac{\tau}{f(n)}\right)\left(1-\frac{\upsilon}{f(n)}\right)},\label{eq:Boseansatz}
\ee
where $\sigma, \tau,\upsilon$ are auxiliary kinematic variables that are some affine transformation of $s,t,u$.  Note that these variables are subject to an auxiliary on-shell condition, $\sigma+\tau+\upsilon=\zeta=$ constant. Here $f(n)$, $g(n)$, and $h(n)$ are unknown functions that diverge as $n\rightarrow \infty$.

Unfortunately, it is far more complicated to solve the polynomial residue constraint for the permutation invariant ansatz in \Eq{eq:Boseansatz} as compared to our earlier cyclic invariant ansatze.  For example, one immediate annoyance is that the auxiliary on-shell condition effectively introduces a new parameter $\zeta$.  Of course, we can instead define $\sigma, \tau,\upsilon$ to be literally equal to $s,t,u$ for massless particles, effectively setting $\zeta=0$.  However, even in this case we find that the solutions for $f,g,h$ satisfying the polynomial residue constraint have zeroes that make the infinite product in \Eq{eq:Boseansatz} ill defined unless we construct the product in a more elaborate way.
Similarly difficulties arise if one attempts to construct permutation invariant amplitudes with bounded spectra starting from an ansatz like \Eq{eq:Boseansatz} but with the kinematic functions inverted to ensure convergence, akin to \Eq{eq:prod}.

For this reason, we leave a detailed analysis of this system for later work.  That said, let us remark on one curious solution to the polynomial residue constraint that we have discovered, which is 
\be
M(s,t) = \frac{(-1)^{\alpha+1}}{stu} \frac{\Gamma(-s)\Gamma(-t)\Gamma(-u)}{\Gamma(s-\alpha)\Gamma(t-\alpha)\Gamma(u-\alpha)} ,\label{eq:VirasoroGen}
\ee
for some nonnegative integer $\alpha$.
For the special case of $\alpha=0$ this is precisely the Virasoro-Shapiro amplitude in \Eq{eq:VS}, while $\alpha>0$, we see that the highest spin exchanged on each pole is finitely increased.  
Here the spectrum of masses is the same as in string theory, in agreement with the no-go result of Ref.~\cite{GeiserUpcoming}.

While this result is enticing, it seems unlikely that the case of general $\alpha$ really offers a consistent deformation of closed-string amplitudes.  In particular, for a bona fide four-point graviton scattering amplitude, one should dress \Eq{eq:VirasoroGen} with the polarization-dependent kinematic function ${\cal R}^4$, which is the multi-linearization of a quartic Riemann operator.  When $\alpha=0$, the massless pole at $s=0$ only encodes spin-two graviton exchange.  However, for $\alpha>0$, this massless pole also includes higher spins, in direct contradiction with Weinberg's theorem.
We leave a thorough analysis of the amplitude in \Eq{eq:VirasoroGen} and its possible cousins for future work.

\section{Discussion}  \label{sec:discussion}

In this paper we have derived a multi-parameter space of Lorentz invariant, four-point, perturbative scattering amplitudes describing the exchange of an infinite tower of higher-spin modes.  By construction, these amplitudes have i) high-energy behavior that is polynomially bounded and ii) poles whose residues encode the exchange of finite-spin degrees of freedom.  Our amplitudes can be viewed as generalizations of the Veneziano and Coon amplitudes, albeit exhibiting modified spin-dependence in their Regge trajectories.

Afterwards, we derived analytic and numerical positivity bounds carving out regions of parameter space in which partial wave unitarity is satisfied.  As far as our analysis is concerned, the remaining islands of positivity are putatively consistent and could in principle correspond to amplitudes of physical theories.

Our work leaves numerous avenues for future inquiry. 
For example, it would be illuminating to obtain a fully analytic understanding of the conditions for partial wave unitarity on our parameter space, $(q, \alpha,\beta,\gamma,\delta)$, as well as the spacetime dimension $D$.  
Another clear direction would be to repeat our analysis with fewer technical assumptions.  For example, our entire setup is based on a product ansatz for the amplitude $A(\sigma,\tau)$.  Also, as we have noted, there may exist alternative solutions to the constraint of polynomial residues, e.g., if we relax the requirement of smoothness on the $f,g,h$ functions or allow $W$ to be a polynomial.  The amplitudes with unbounded spectra described in \Sec{sec:variations} are also worthy of further study.

In terms of the underlying physics, our most pressing concern is to what extent the amplitudes  we have derived here can be generalized to higher point.  Indeed, the fact that there is an abundance of four-particle amplitudes satisfying conditions i) and ii) might simply be a sign that constraints on four-particle scattering are simply {\it not that strong}.  If this is the case, then deformations of higher-point string scattering will be pathological.  It should be noted that there exists prior work constructing higher-point generalizations of the Coon amplitude~\cite{Baker:1970vxk,Fairlie:1994ad}.

Finally, it would be interesting to understand if there is a more direct physical interpretation of the Coon amplitude and our generalizations thereof.  Indeed the spectrum is remarkably rich: a discretum of hydrogen-like levels culminating at an accumulation point, above which lies a branch cut suggesting a continuum of ionized states. While string theory realizes amplitudes with features matching the Coon form~\cite{Maldacena:2022ckr}, a more general physical understanding is warranted.

\vspace{\baselineskip}

\begin{center} 
{\bf Acknowledgments}
\end{center}
\noindent 
We would like to thank Nima Arkani-Hamed, Lorenz Eberhardt, Nick Geiser, Andreas Helset, Yu-tin Huang, Juan Maldacena, Sebastian Mizera, Julio Parra-Martinez, John Schwarz, and Piotr Tourkine for useful comments.
C.C. is supported by the Department of Energy (Grant~No.~DE-SC0011632) and by the Walter Burke Institute for Theoretical Physics. 
G.N.R. is supported at the Kavli Institute for Theoretical Physics by the Simons Foundation (Grant~No.~216179) and the National Science Foundation (Grant~No.~NSF PHY-1748958) and at the University of California, Santa Barbara by the Fundamental Physics Fellowship.

\vspace{\baselineskip}

\appendix
\section{Analytic Partial Waves}\label{app:partials}
Consider the $(\alpha,\beta,\gamma)$ amplitude defined in Eq.~\eqref{eq:Afull}. 
Recall that, as in the case of the Coon amplitude~\cite{Tourkine}, we assume that the external particles are scalars whose masses are the same as the lowest state exchanged in the amplitude, which in our notation is at $\mu^{2}\delta=m_{0}^{2}$. 
The residue on the pole at $s=m_{k}^{2}=\mu^{2}\left(\delta+\frac{q^{k}-1}{q-1}\right)$
in Eq.~\eqref{eq:Afull} is given by Eq.~\eqref{eq:Rkproduct}.
By construction, $R_{k}$ is a finite polynomial in $\tau$.
Written in terms $x=\cos\theta$ for $\theta$ the physical scattering angle defined in Eq.~\eqref{eq:costheta}, we have the pole at $\sigma=q^{k}$,
\be 
\begin{aligned}
R_{k} & =c_{\alpha\beta\gamma} q^{k(\alpha+\beta)}\prod_{j,j\neq k}\left(1-q^{j-k}\right)^{-1}\times\\
 & \qquad\qquad\times\left[1+\frac{1}{2}\left(1-q^{k}\right)(1-x)+\delta\frac{1-q}{2}(3x-1)\right]^{\beta+\gamma k}\times\\
 & \qquad\qquad\times\prod_{l=0}^{k+\alpha-1}\left[1+\frac{1}{2}\left(1-q^{k}\right)(1-x)+\delta\frac{1-q}{2}(3x-1)-q^{l-k-\alpha}\right].
\end{aligned}
\label{eq:Rk}
\ee 
We can write this expression in terms of the parameters in Eq.~\eqref{eq:AB} as
\be 
\begin{aligned}
R_{k}&=\frac{c_{\alpha\beta\gamma}q^{k(\alpha+\beta)}}{(q;q)_\infty (q^{-k};q)_k}(-b_{k})^{\ell_{\max}(k)}(x-x_{\infty,k})^{\beta+\gamma k}\prod_{l=0}^{k+\alpha-1}(x-x_{l,k}) \\
&=\frac{c_{\alpha\beta\gamma}q^{k(\alpha+\beta)}}{(q;q)_{\infty}(q^{-k};q)_{k}}\left(-b_k(x-x_{\infty,k})\right)^{\ell_{\max}(k)}\left(-\frac{1}{q^{k+\alpha}b_k (x-x_{\infty,k})};q\right)_{k+\alpha},
\end{aligned}
\end{equation}
where $\ell_{\max}(k)$ is defined in Eq.~\eqref{eq:lmax}. Using Eq.~\eqref{eq:qbinomial} to rewrite the $q$-Pochhammer symbol in terms of $q$-binomial coefficients, we have
\begin{equation}
R_{k}=\frac{c_{\alpha\beta\gamma}q^{k(\alpha+\beta)}}{(q;q)_{\infty}(q^{-k};q)_{k}}\left(-b_k (x{-}x_{\infty,k})\right)^{\ell_{\max}(k)}\sum_{j=0}^{k+\alpha}q^{-j(k+\alpha)}\left(b_k(x{-}x_{\infty,k})\right)^{-j}q^{\frac{j(j-1)}{2}}\genfrac[]{0pt}{1}{k+\alpha}{j}_q,\label{eq:Rksum}
\end{equation}
or equivalently, relabeling $j\rightarrow k+\alpha-j$,
\begin{equation}
R_{k}=\frac{(-1)^{\ell_{\max}(k)}c_{\alpha\beta\gamma}q^{k(\alpha+\beta)}}{(q;q)_{\infty}(q^{-k};q)_{k}}\sum_{j=0}^{k+\alpha}\left(b_k(x{-}x_{\infty,k})\right)^{\beta+\gamma k + j}q^{-\frac{(k+\alpha-j)(k+\alpha+j+1)}{2}}\genfrac[]{0pt}{1}{k+\alpha}{k+\alpha-j}_q.
\end{equation}
Using the binomial theorem, we can rewrite this expression as
\begin{equation}
\begin{aligned}R_{k} & =(-1)^{\alpha+\beta+(1+\gamma) k}\frac{c_{\alpha\beta\gamma}q^{k(\alpha+\beta)}}{(q;q)_{\infty}(q^{-k};q)_{k}}
\times \\
&\times \left(\sum_{j=0}^{k+\alpha}(b_{k})^{j+\beta+\gamma k}q^{-\frac{(k+\alpha-j)(k+\alpha+j+1)}{2}}\genfrac[]{0pt}{1}{k+\alpha}{k+\alpha-j}_q \sum_{r=0}^{j+\beta+\gamma k}\tfrac{(j+\beta+\gamma k)!}{r!(j+\beta+\gamma k-r)!}(-x_{\infty,k})^{j+\beta+\gamma k-r}x^{r}\right).
\end{aligned}
\end{equation}
Now, we can rearrange a general double sum via the identity
\begin{equation}
\sum_{j=0}^{n}\sum_{r=0}^{j+N}M_{r,j}x^{r}=\sum_{s=0}^{N}\sum_{j=0}^{n}M_{s,j}x^{s} +\sum_{p=1}^{n}\sum_{j=p}^{n}M_{N+p,j}x^{N+p}.
\end{equation}
Thus, rearranging the sums to group all terms going like $x^{s}$
together, we have
\begin{equation}
R_{k} =(-1)^{\alpha+\beta+(1+\gamma)k}\frac{c_{\alpha\beta\gamma}q^{k(\alpha+\beta)}}{(q;q)_{\infty}(q^{-k};q)_{k}}\left(\sum_{s=0}^{\beta+\gamma k}\sum_{j=0}^{k+\alpha}+\sum_{s=\beta+\gamma k+1}^{\alpha+\beta+(1+\gamma)k}\sum_{j=s-\beta-\gamma k}^{k+\alpha}\right)Q_{j,s},
\end{equation}
where 
\begin{equation}
Q_{j,s}=(b_{k})^{j+\beta+\gamma k}q^{-\frac{(k+\alpha-j)(k+\alpha+j+1)}{2}}\genfrac[]{0pt}{1}{k+\alpha}{k+\alpha-j}_q\,\tfrac{(j+\beta+\gamma k)!}{s!(j+\beta+\gamma k-s)!}(-x_{\infty,k})^{j+\beta+\gamma k-s}x^{s}.
\end{equation}
Now, the monomial $x^{r}$ can be written in terms of the Gegenbauer
polynomials,
\begin{equation}
x^{r}=\sum_{v=0}^{\lfloor r/2\rfloor}\frac{r!\left(1+\frac{2(r-2v)}{D-3}\right)\Gamma\left(\frac{D-1}{2}\right)}{2^{r}v!\Gamma\left(\frac{D-1}{2}+r-v\right)}G_{r-2v}^{(D)}(x).\label{eq:monomialGegenbauer}
\end{equation}
Expanding $R_k$ in partial waves as in Eq.~\eqref{eq:Rkpartials}, we therefore have
\begin{equation}
a_{k,\ell}=\sum_{s=0}^{\beta+\gamma k}\sum_{\nu=0}^{\lfloor s/2\rfloor}\sum_{j=0}^{k+\alpha}Q_{j,\nu,s}\delta_{\ell,s-2\nu}+\sum_{s=\beta+\gamma k+1}^{\alpha+\beta+(1+\gamma)k}\sum_{\nu=0}^{\lfloor s/2\rfloor}\sum_{j=s-\beta-\gamma k}^{k+\alpha}Q_{j,\nu,s}\delta_{\ell,s-2\nu},
\end{equation}
where
\begin{equation}
\begin{aligned}Q_{j,\nu,s} & =\left(1+\frac{2\ell}{D-3}\right)\frac{\Gamma\left(\frac{D-1}{2}\right)}{2^{s}v!\Gamma\left(\frac{D-1}{2}+s-v\right)}(-1)^{\alpha+\beta+(1+\gamma)k}\frac{c_{\alpha\beta\gamma}q^{k(\alpha+\beta)}}{(q;q)_{\infty}(q^{-k};q)_{k}}\times\\
 & \;\;\;\;\times(b_{k})^{j+\beta+\gamma k}q^{-\frac{(k+\alpha-j)(k+\alpha+j+1)}{2}} \genfrac[]{0pt}{1}{k+\alpha}{k+\alpha-j}_q\frac{(j+\beta+\gamma k)!}{(j+\beta+\gamma k-s)!}(-x_{\infty,k})^{j+\beta+\gamma k-s}.
\end{aligned}
\end{equation}
Now, one can check that, for any matrix $M_{v,s}$, the following identities hold:
\be 
\begin{aligned}
\sum_{s=0}^{N}\sum_{v=0}^{\lfloor s/2\rfloor}M_{v,s}\delta_{\ell,s-2v}&=\sum_{s=0}^{\lfloor(N-\ell)/2\rfloor}M_{s,\ell+2s}\\
\sum_{s=N+1}^{N+n}\sum_{v=0}^{\lfloor s/2\rfloor}M_{v,s}\delta_{\ell,s-2v}&=\sum_{s=\max\left\{0,1{+}\left\lfloor(N-\ell)/2 \right\rfloor \right\}}^{\lfloor(N+n-\ell)/2\rfloor}M_{s,\ell+2s}.
\end{aligned}
\ee
Putting this all together, we obtain
\be
a_{k,\ell}=\sum_{s=0}^{\lfloor\frac{\beta+\gamma k-\ell}{2}\rfloor}\sum_{j=0}^{k+\alpha}Q_{j,s,\ell+2s} + \sum_{s{=}\max\left\{0,1{+}\left\lfloor \frac{\beta{+}\gamma k{-}\ell}{2}\!\right\rfloor \!\right\}}^{\lfloor\frac{\alpha+\beta+(1+\gamma)k-\ell}{2}\rfloor}\sum_{j=\ell+2s-\beta-\gamma k}^{k+\alpha}Q_{j,s,\ell+2s},
\ee
which is the closed-form expression for all partial wave coefficients given in \Eq{eq:aa}.

\bibliographystyle{utphys-modified}
\bibliography{ProductAmps}

\end{document}